\documentclass{SciPost}

\hypersetup{
	colorlinks,
	linkcolor={scipostblue},
	citecolor={blue!50!black},
	urlcolor={scipostblue}
}

\usepackage[bitstream-charter]{mathdesign}
\usepackage{bm}
\usepackage{graphicx}
\usepackage{tikz}
\usepackage{genyoungtabtikz}
\usetikzlibrary{calc}

\newcommand{\cO}{\mathcal{O}}
\newcommand{\cI}{\mathcal{I}}
\newcommand{\cN}{\mathcal{N}}
\newcommand{\cA}{\mathcal{A}}

\newcommand{\dota}{\dot{\alpha}}
\newcommand{\dotb}{\dot{\beta}}
\newcommand{\dotc}{\dot{\gamma}}
\newcommand{\fa}{\mathfrak{a}}
\newcommand{\fb}{\mathfrak{b}}
\newcommand{\dfa}{\dot{\mathfrak{a}}}
\newcommand{\dfb}{\dot{\mathfrak{b}}}
\newcommand{\bb}{\beta}
\newcommand{\hX}{\hat{X}}

\newcommand{\bX}{\bar{X}}
\newcommand{\GL}{\mathrm{GL}}
\newcommand{\SL}{\mathrm{SL}}
\newcommand{\SU}{\mathrm{SU}}

\newcommand{\ula}{\underline{\lambda}}
\newcommand{\uLa}{\underline{\Lambda}}
\newcommand{\uGa}{\underline{\Gamma}}
\newcommand{\umu}{\underline{\mu}}
\newcommand{\unu}{\underline{\nu}}
\newcommand{\SO}{\mathrm{SO}}
\newcommand{\U}{\mathrm{U}}

\makeatletter
\newcommand{\llangle}{\mathopen{\langle\mkern-4mu\langle}}
\newcommand{\rrangle}{\mathclose{\rangle\mkern-4mu\rangle}}
\makeatother

\DeclareSymbolFont{usualmathcal}{OMS}{cmsy}{m}{n}
\DeclareSymbolFontAlphabet{\mathcal}{usualmathcal}

\begin{document}
	
	\begin{center}
	    {}
	\end{center}
    \begin{center}
	    {}
	\end{center}
	\begin{center}{\Large  \textbf{\color{black}{
					A Compact Formula for Conserved Three-Point Tensor Structures in 4D CFT
	}}}\end{center}
	
	\begin{center}\textbf{
			Paul Heslop and Hector Puerta Ramisa
		}
	\end{center}
	
	\begin{center}
		Department of Mathematical Sciences, Durham University, Durham, UK
		\\[\baselineskip]
		\href{mailto:email1}{\small paul.heslop@durham.ac.uk}\,,\ \href{mailto:email1}{\small hector.puerta-ramisa@durham.ac.uk}
	\end{center}
	
	\vspace{\baselineskip}

	\section*{\color{black}{Abstract}}
	\textbf{\boldmath{%
			We 
			derive a compact analytic formula for a complete basis of conformally invariant tensor structures for three-point functions of conserved operators in arbitrary 4D Lorentz representations.
			The construction follows directly from a novel constraint  equivalent to applying conservation conditions at each point:  the leading terms  in all OPE limits appear as  symmetric traceless tensors. 
			We derive this by lifting  to a unified $\SU(m,m|2n)$ analytic superspace framework, where the conservation conditions are automatically solved
			and then reducing back to 
			4D CFT.
			The same method is also used for cases involving one non-conserved operator.
			This formalism further reveals a map of the counting of CFT tensor structures to that of finite-dimensional
			$\SU(2n)$ representations, solved by Littlewood-Richardson coefficients.
			All results can be directly re-interpreted as three-point $\cN=2$ and $\cN=4$ superconformal tensor structures via the unified analytic superspace.
	}}
	
	\vspace{\baselineskip}

	\vspace{10pt}
	\noindent\rule{\textwidth}{1pt}
	\newpage
	{\hypersetup{
			linkcolor={black}
		}
		\tableofcontents}
	\noindent\rule{\textwidth}{1pt}
	\vspace{10pt}

	\newpage 
	\section{Introduction}
	
	In recent years, the conformal bootstrap has become a central tool for the study of a wide array of physical systems from condensed matter to quantum gravity.
	This programme aims to  completely determine,  or at least vastly restrict,   the allowed space of conformal field theories (CFT) at the fully non-perturbative level using fundamental consistency arguments. 
	The high degree of symmetry of these theories imposes strong constraints on their dynamics: they are uniquely defined by 
	the spectrum of operators---most notably their conformal dimensions---and the three-point (OPE) coefficients. 
	The central idea of the conformal bootstrap is to extract this data by enforcing the compatibility of crossing symmetry of four-point functions with the operator product expansion.  
	The vast majority of work in this area so far has involved studying scalar four-point correlators, and the related  three-point functions of two scalars and one non-scalar operator.

		Going beyond the scalar case is crucial for further progress but many complications appear. 
		The very first complication is that  the three-point functions no longer have a unique functional form, but instead there can be multiple independent tensor structures consistent with conformal symmetry. The CFT data must then incorporate coefficients for each of these tensor structures. Thus, the very first thing to understand when going beyond scalar correlators is this space of tensor structures. There has been  steady progress in understanding this for over fifty years,  e.g.
		\cite{Schreier:1971um,Osborn:1993cr,Petkou:1994ad,Costa:2011mg,Stanev:2012nq,Elkhidir:2014woa,Cuomo:2017wme,Buchbinder:2023coi, Baumann:2024ttn}.
		
		Three-point functions of conserved currents (arbitrary Lorentz rep) are a particularly important class of CFT observables. Indeed, beyond fundamental fields, conserved currents are the most basic conformal multiplets, saturating the unitarity bound and therefore possessing the lowest possible twist.
		They are of immediate relevance in holographic contexts, where they are dual to massless higher-spin fields in the bulk, and thus their CFT three-point functions at the boundary encode the different possible interactions in the bulk.
		Fixing three-point tensor structures of such fields is therefore of direct relevance to higher-spin holography \cite{Vasiliev:1990en,Giombi:2009wh,Didenko:2014dwa, Boulanger:2015ova, Sleight:2016dba}.
		They similarly probe de Sitter physics, making them relevant to holographic approaches to cosmology from 3D CFT~\cite{Strominger:2001pn, McFadden:2009fg,Maldacena:2011nz,McFadden:2011kk,Arkani-Hamed:2015bza,Baumann:2020dch}.

		Despite their simplicity as conformal multiplets, their three-point tensor structures are harder to fix than those of generic (unconstrained) spinning operators due to the additional conservation constraint which needs to be imposed.
		In doing so, the space of tensor structures for the three-point correlator gets reduced. The first example of this was given in \cite{Schreier:1971um,Osborn:1993cr} for three-point functions of currents and stress-tensors, respectively. For arbitrary spin symmetric traceless tensors, the number of independent structures before imposing conservation scales with the product of their spins (e.g.~\eqref{eq:ssscounting}). Whereas after imposing conservation at each point, this number reduces to $ 2 \min(s_1,s_2,s_3) +1$~\cite{Costa:2011mg,Stanev:2012nq}.
		More recently, three-point functions of two mixed-symmetry representations and a third symmetric traceless insertion have been studied in \cite{Buchbinder:2023coi}, arriving at a corresponding counting.
		We note that the CFTs4D \textsf{Mathematica} package \cite{Cuomo:2017wme} allows for explicit checks involving correlators of low-spin fields, but the counting has remained conjectural until now. 
		
		Recent interest in this problem has been driven by new approaches using twistor theory to solve conservation constraints~\cite{Baumann:2024ttn,CarrilloGonzalez:2025qjk}.
		The main advantage of these constructions is that conservation conditions are automatically satisfied on twistor space by choosing appropriate holomorphicity requirements.
		However, the map back to position space then takes the form of a nested integral transform which needs to be evaluated on a case-by-case basis.         
        
        Another interesting development in recent years has arisen from the observation~\cite{Hofman:2008ar,Costa:2011mg,Kravchuk:2016qvl,Costa:2016hju} that the number of  structures for CFT correlators in $d$ dimensions equals the number of structures for QFT scattering amplitudes in $d+1$ dimensions. This has led to interesting work seeking a direct map between the two structures~\cite{Caron-Huot:2021kjy,Lee:2023qqx}.
		
		Despite all of this progress,  there has until now been no concrete analytic formula describing the space of general conserved three-point functions. One of the main outputs of this paper is to fill this gap for 4D CFT, providing precisely such an  explicit  formula which in fact turns out to be remarkably simple. We also prove and  generalise the counting formulae above. Although the focus is on 4D, we expect all our methods including the OPE limit statements to apply naturally to CFT in other dimensions $d\geq3$.
		
		Our approach,  similarly to twistor theory, considers a space where conservation is automatic, realised by a lift to analytic superspace~\cite{Galperin:1984av,Howe:1995md,Heslop:2001gp,Heslop:2003xu,Heslop:2022xgp,Hansen:2025lig}.  Despite the fact that we are considering non-supersymmetric theories,  viewing 4D Minkowski case as a special case of analytic superspace is  a very fruitful idea. The point is that all three-point structures have direct uplifts to analytic superspace, but in that context the conservation condition is automatic, arising from analyticity in certain internal variables. 
		The resulting structures can then be directly reduced back down to  Minkowski space.
		
		The result of this analysis can be phrased  purely  in 4D position space,  without introducing the full analytic superspace  machinery. 
		The key constraint gives the form of the tensor structures in coincident limits. This constraint is equivalent to  conservation conditions at each point, but is simpler to solve and leads directly to our analytic formula.

		The paper proceeds as follows. In section~\ref{sec:4D}, we set up the problem in purely 4D CFT terms using Weyl spinor indices.  We review the space of tensor structures and the identities between them in the absence conservation constraints.
		The main results of our paper are presented in section~\ref{sec:results}. We state (proved in section \ref{sec:coset}) the main constraint on coincident limits and solve it to arrive at our main formula~\eqref{main}: an explicit basis for three-point tensor structures of conserved operators with total left and right spin equal. 
		We also discuss cases involving a non-conserved insertion in the OPE of two conserved tensors.
		In section~\ref{sec:coset}, we define and construct the unified $\SU(m,m|2n)$ analytic superspace and use it to prove the validity of the  constraint we imposed on the conserved non-supersymmetric tensor structures. 
		We also point out an application of the universality of analytic superspace, which is that counting of conformally-invariant tensor structures (both conserved and not) can be performed using purely finite-dimensional $\SU(2n)$ representation theory using Littlewood-Richardson coefficients. 
		Finally, we conclude and provide directions of future work in section~\ref{conclusion}. 
		We also include two appendices outlining specific checks (appendix \ref{app:CF}) and the theory and computation of Littlewood-Richardson coefficients (appendix \ref{app:LR}).

		\section{Conformal Correlators in 4D}
		\label{sec:4D}

		We wish to count and write down an explicit  basis for general 3-point correlators in 4D, in particular focussing on the cases where some or all operators are conserved. 
		The functional form of two-point functions of any two operators is unique, as is that of three-point functions involving at least two scalar operators. This is no longer true in general for three-point functions where there are two or more non-scalar operators.  
		
		In this section we will set up the problem, emphasising the usefulness of Weyl spinor notation in 4D, where solving the Ward identities is equivalent to tracking indices locally at each point.  We write down the solution for non conserved operators including an analysis of the  identities which arise from 4D kinematics.

		\subsection{Conformal symmetry of correlation functions}

		Local operators in a CFT are representations of the conformal group, specified by their Lorentz representation, together with their conformal dimension $\Delta$. In 4D the massless Lorentz rep is determined  by the left and right spin $s,\bar s$ and the corresponding operator 
		has $s$ symmetrised undotted Weyl spinor indices and $\bar s$ dotted ones:
		\begin{align}
			\cO_{s,\bar s}^\tau = \cO^\tau_{\underbrace{\scriptstyle (\alpha\beta\dots \gamma)}_{\textstyle s},\underbrace{\scriptstyle (\dot \alpha \dot \beta,\dots \dot \gamma)}_{\textstyle \bar s}}\,. 
		\end{align}
		Here we label by the twist $\tau$ rather than conformal dimension, given by
		\begin{equation}\label{eq:twist}
			\tau=\Delta-\tfrac12(s{+}\bar s).
		\end{equation}
		Then correlators, vacuum expectation values of local operators, are tensor functions of space-time points $x_{i=1,\dots,n}$ which it is convenient to view in spinor notation $x^{\alpha \dot \alpha}= x^\mu \sigma_\mu^{\alpha \dot \alpha}$. The tensor function has the indices of all $n$ operators.
		So for example for a correlator of $n$ spin 1 operators: 
		\begin{align}
			\langle \cO^{\tau_1}_{\alpha_1 \dot \alpha_1}(x_1)\dots \cO^{\tau_n}_{\alpha_n \dot \alpha_n}(x_n) \rangle = T_{\alpha_1\dot \alpha_1\dots \alpha_n \dot \alpha_n}(x_i)\,.
		\end{align}
		The tensor function $T$ then transforms under conformal transformations at each point in the same way as each operator. The consequences of this are in fact very  simple to describe in 4d in Weyl spinor notation. Essentially build a function from differences of space-time points $x_{ij}=x_i-x_j$,  inverses of these as a 2$\times$2 matrix 
		\begin{equation}\label{eq:Iij}
			(I_{ji})_{\dota \alpha} =(x_{ij}^{\alpha \dota})^{-1}
		\end{equation}
		together with $x^2_{ij}$.%
        \footnote{We here assume the total left and right spins of all operators in the correlator are equal. If this is not the case a further ingredient is needed, see section~\ref{unbalanced}. } 
        All indices have to be contracted `locally' (meaning only indices associated with the same space-time point can be contracted) apart from remaining uncontracted indices corresponding to the indices of the correlator. Then the powers of $x_{ij}^2$ have to be such that the weight of point $x_i$ is $-\tau_i$ to match the twist of the operator at point $i$. (The contributions from  $x_{ij}$ and $I_{ij}$ to the twist vanish.)

		For example two-point functions have the unique functional form:
		\begin{equation}
			\langle \cO^\tau_{\underbrace{\scriptstyle (\alpha_1\dots \gamma_1)}_{\textstyle s},\underbrace{\scriptstyle( \dot \alpha_1 \dots \dot \gamma_1)}_{\textstyle\bar s}} \cO^\tau_{\underbrace{\scriptstyle (\alpha_2\dots \gamma_2)}_{ \textstyle \bar s},\underbrace{\scriptstyle (\dot \alpha_2  \dots \dot \gamma_2)}_{\textstyle s}}\rangle = \frac{N}{|x_{12}|^{2\tau}}(I_{12})_{(\dota_1 (\alpha_2}\dots (I_{12})_{\dot \gamma_1)\gamma_2)}
			(I_{21})_{(\dota_2 (\alpha_1}\dots (I_{21})_{\dot \gamma_2)\gamma_1)}
		\end{equation}
		The RHS can only be written down using the above rules if  the two operators are in conjugate representations $(\cO^\tau_{s,\bar{s}})^\dagger = \cO^\tau_{\bar{s},s}$, and the two-point function vanishes otherwise.

		We write this in the simplified notation with the indices suppressed
		\begin{equation}\label{eq:2ptcorr}
			\langle \cO_{s, \bar{s}}^\tau(x_1) \cO^\tau_{\bar{s},s}(x_2)\rangle = N\frac{I^{s}_{12} I_{21}^{\bar{s}}}{|x_{12}|^{2\tau}}\,.
		\end{equation}
		The indices can be simply and uniquely restored, since their positions are dictated by the space-time points. 
		Alternatively one can view  the open indices to be contracted by polarisation spinors, $s_i^\alpha,\bar s_i^{\dot \alpha}$, as is commonly done in this context, so that 
		\begin{equation}
			\cO_{s,\bar s}^\tau\equiv \cO^\tau_{
				\alpha\beta\dots \gamma, \dot \alpha \dots \dot \gamma} (s^\alpha \dots s^\gamma) (\bar s^{\dot \alpha} \dots \bar s^{\dot \gamma})\,,
		\end{equation}
		and 
		\begin{equation}
			I_{12}^s \equiv (\bar{s}_1^{\dota} I_{12 \dota \alpha} s_2^{\alpha})^s.
		\end{equation}

		\subsection{Non-conserved tensor structures}\label{sec:corr}

		For  three-point function of spinning fields we label the different possible tensor structures with an index $(i)$ and distinguish them from the correlator by double brackets $\llangle \dots \rrangle$ as follows
		\begin{equation}
			\langle \cO_1 \cO_2 \cO_3\rangle = \sum_{i}\lambda_{123}^{(i)} \llangle  \cO_1 \cO_2 \cO_3 \rrangle^{(i)}\,.
		\end{equation}
		We consider correlators of operators with arbitrary left and right spins, but focus here on the case for which the total left spin equals the total right spin   
		\begin{equation}\label{eq:balanced}
			\sum_i^3 s_i - \bar{s}_i=0\, ,
		\end{equation}
		which we call balanced correlators.
		We discuss the generalisation beyond this restriction in section~\ref{unbalanced}.
		
		Three-point structures of balanced correlators are combinations of $I_{ij}$ \eqref{eq:Iij} and the following basic covariant
		\begin{equation}\label{eq:covariants}
			(Y_{i,jk})_{\dota \alpha } = (I_{ij}x_{jk}I_{ki})_{\dota \alpha } = (x_{ji}^{-1}-x_{ki}^{-1})_{\dota\alpha},
		\end{equation}
		with no further contraction of indices.
		Note $Y_{i,jk}=-Y_{i,kj}$ so we identify $Y_i$ with the positive permutation of $(ijk)$ in $S_3$.
		This set of basic covariants can be neatly packed into a $3\!\times\!3$ matrix $V_{ij}$ defined as 
		\begin{align}\label{Vs}
			\begin{pmatrix}V_{11}&V_{12}&V_{13}\\V_{21}&V_{22}&V_{23}\\V_{31}&V_{32}&V_{33}\end{pmatrix} := \begin{pmatrix} Y_{1}&I_{12}&I_{13}\\-I_{21}&Y_{2}&I_{23}\\-I_{31}&-I_{32} &Y_{3}\end{pmatrix} .
		\end{align}
		Then, a three-point structure of generic fields $\cO^{\tau_i}_{s_i,\bar{s}_i}$ with twist $\tau_i$ and in Lorentz rep $[s_i,\bar{s_i}]$ takes the following form
		\begin{equation}\label{eq:genstructure}
			\llangle \cO^{\tau_1}_{s_1,\bar{s}_1}\cO^{\tau_2}_{s_2,\bar{s}_2}\cO^{\tau_3}_{s_3,\bar{s}_3}\rrangle=P \prod_{i,j=1}^3 V_{ij}^{ b_{ij}},
		\end{equation}
		where $P$ is the scalar prefactor
		\begin{equation}\label{eq:K3}
			P= 
			\frac{1}{(x_{12}^2)^{\frac{\tau_1+\tau_2-\tau_3}2}(x_{23}^2)^{\frac{\tau_2+\tau_3-\tau_1}2}(x_{31}^2)^{\frac{\tau_3+\tau_1-\tau_2}2}},
		\end{equation}
		where $b_{ij}$ are positive integers satisfying
		\begin{equation}
			\sum_{i=1}^3b_{ij} = s_j, \qquad \sum_{j=1}^3 b_{ij} = \bar{s}_i \, .
		\end{equation}
		The number of such structures is equivalent to the number of non-negative integer $3\times 3$ matrices with rows summing to $s_i$ and columns to $\bar s_j$ (known as a contingency table). For equal spin $s_i=\bar s_i=s$, the number of these is 
		\begin{align}\label{binom}
			\binom{s+5}{5}-\binom{s+2}{5}\,.
		\end{align}
		There is no closed formula for the counting for general spins, but there is a useful generating function for it:
		\begin{align}\label{genfun}
			\left.\prod_{i,j=1}^3\frac1{1-x_iy_j}\right|_{x_i^{s_i}y_j^{\bar s_j}}\,.
		\end{align}

		\subsection{Identities}\label{sec:identities}
		The  three-point structures~\eqref{eq:genstructure} are not independent in 4D since there are  identities between them, which arise from antisymmetrising more than two Weyl spinor indices.
		The easiest way to see this is to use a conformal transformation to set $x_3\rightarrow \infty$ and $x_1 \rightarrow 0$, then we let $x_2=x$. To avoid infinities, we multiply any spinor indices associated with point 3, $\dot \alpha$ by $(I_{12}x_{23})_{\dot \beta}{}^{\dot \alpha} $ and $\alpha$ by $(x_{32}I_{21})^{\alpha}{}_{\beta} $. 
		In this conformal frame we have that
		\begin{align}\label{cframe}
			Y_i,I_{12},I_{23},I_{13} \rightarrow  x^{-1}, \qquad   I_{31},I_{32},I_{21} \rightarrow - x^{-1}\,.
		\end{align}
		Then, we see that any antisymmetric combination of the elements $V_{ij}$~\eqref{Vs} 
		will vanish, since in this conformal frame it corresponds to antisymmetrising the indices of $(X^{-1})_{\dot \alpha \alpha}$.
		
		Indeed, we see that all identities in these three-point functions reduce down to the simplest one which  occurs for the  correlator of spin-1 fields and is given (with indices following the particle numbering) by \begin{align}\label{idV}
			\cI \equiv V_{1i}V_{2j}V_{3k} \epsilon^{ijk}=0.
		\end{align}
		One sees this must vanish by going to the above conformal frame, where it  corresponds to explicitly antisymmetrising three 2-component Weyl spinor indices: 
		$$(x^{-1})_{\dot \alpha_1[\alpha_1}(x^{-1})_{|\dot \alpha_2|\alpha_2}(x^{-1})_{|\dot \alpha_3|\alpha_3]} =0.$$
		Writing out the $V$s in terms of $Y$'s and $I$'s according to~\eqref{Vs}, this becomes
		\begin{equation}\label{eq:2}
			{\mathcal I}_{\alpha_1\alpha_2\alpha_3\dot\alpha_1\dot\alpha_2\dot\alpha_3} \equiv 
			Y_1 Y_2 Y_3 + Y_1 I_{23}I_{32}+Y_2 I_{13}I_{31}+Y_3 I_{12}I_{21}- I_{12}I_{23}I_{31}+I_{13}I_{32}I_{21}=0\, ,
		\end{equation}
		where on the right hand side we have again suppressed the indices, but they follow the particle number, e.g. $Y_i$ really means $(Y_i)_{\dot \alpha_i \alpha_i}$  and $I_{ij}$ means $(I_{ij})_{\dot \alpha_i \alpha_i}$ as discussed below~\eqref{eq:Iij}. 
		
		All other higher-spin three-point identities can be similarly obtained antisymmetrising 3 or more $V$s. In fact,  they are all generated by multiplying this simple identity by arbitrary structures as follows 
		\begin{align}\label{eq:HSids}
			\prod_{i,j=1}^3V_{ij}^{b'_{ij}} \times {\mathcal I} =0,   \qquad  \sum_i b'_{ij} = s_i-1, \qquad \sum_j b'_{ij} = \bar{s}_j-1  \, .
		\end{align}
		Thus  the number of these identities is equal to the number of
		three point structures (without identities) with spins $s_i-1$ and $s_j-1$.
		Thus following~\eqref{genfun}, the number of independent terms in the general case is given by the generating function
		\begin{align}\label{noind3}
			\left.
			\frac{1-x_1x_2x_3y_1y_2y_3}{\prod_{i,j=1}^3(1-x_iy_j)}\right|_{x_i^{s_i}y_j^{\bar s_j}}\,.
		\end{align}
		A basis of independent cases can be found by simply removing any term containing one of the factors in~\eqref{eq:2}, for example excluding any term containing a factor of $I_{13}I_{32}I_{21}$.
		
		For all equal spins $s_i=\bar s_i=s$ this counting reduces from~\eqref{binom} to 
		\begin{align}\label{eq:ssscounting}
			\binom{s+5}{5}-\binom{s+2}{5}- \binom{s+4}{5}+\binom{s+1}{5}=\frac12 (1+s)(2+2s+s^2)\,.
		\end{align}
		For certain more general subsets of the general case (in particular, when at least two  operators have equal left and right spin) analytic formulae have been found for the counting~\cite{Costa:2011mg,Elkhidir:2014woa} which agree with this generating function.

		\subsection{Correlators with spin imbalance}
		\label{unbalanced}

		It is convenient to characterise three-point functions in 4D by the difference between undotted and dotted spinor indices, which must be an even number:
		\begin{equation}\label{eq:4Dbalance}
			2B= \sum_i^3 s_i - \sum_i^3 \bar{s}_i \ ,\qquad B\in \mathbb{Z}.
		\end{equation}
		We call  $B$ the `imbalance' of the three-point function. A correlator is then called balanced if $B=0$ and unbalanced otherwise.

		We have focussed so far on balanced correlators. Three-point functions with non-zero imbalance \eqref{eq:4Dbalance} are constructed with the additional covariant
		\begin{equation}\label{eq:KK}
			(K_{ij,k})^{\alpha}_{\ \beta} := x_{ik}^{\alpha \dota}(I_{kj})_{\dota \beta}, \quad (B>0)\qquad \qquad (\bar{K}_{ij,k})_{\dota}^{\ \dotb} := (I_{ik})_{\dota \alpha} x_{kj}^{\alpha \dotb},\quad (B<0)
		\end{equation}
		which adds or subtracts to the imbalance $B$.
		Indeed, unbalanced tensor structures take the form of a balanced piece appropriately dressed with the above covariants.
		The general approach is as in the balanced case, namely write the most general product of covariants $I_{ij}$, $Y_{i}$, $K_{ij,k}$, $\bar{K}_{ij,k}$ and identify identities thus reducing to a basis of linearly independent structures. 
		For this, one can find all possible shortest ways to reduce from unbalanced correlators to balanced ones by using $K$ and $\bar{K}$, similarly to weight-shifting operators \cite{Karateev:2017jgd,Hansen:2025lig}, and then impose the identities.
		This generalised exercise was done in \cite{Elkhidir:2014woa}. 
		We note that, as in the previous section, the identities follow from antisymmetrising 3 Weyl spinor indices.

			\section{Conserved three-point structures}\label{sec:results}

			In the previous section, we constructed a basis of conformally invariant three-point structures for generic, non-conserved operators. However, when one or more operators have sufficiently low twist \eqref{eq:twist}, they satisfy additional (conservation) constraints which need to be satisfied by correlators.
			
			In particular, a conserved field in a four-dimensional conformal field theory is a tensor $J$ in the $(s,\bar{s})$ representation of the Lorentz group with conformal dimension saturating the unitarity bound
			\begin{equation}\label{eq:unitarity bound}
				\Delta\geq2+\frac{s+\bar{s}}{2},
			\end{equation}
			so twist  $\tau=2$.
			When this bound is saturated, certain descendants have vanishing norms, and thus must themselves vanish by unitarity. In particular, 
			\begin{equation}\label{eq:constraint}
				\partial^{\alpha_1 \dota_1} J_{\alpha_1 \dots \alpha_s, \dota_1 \dots \dota_{\bar{s}}} =0\,.
			\end{equation}
			These are additional constraints that must be imposed on the structures \eqref{eq:genstructure}, thus reducing the basis. In this section, we count and write down a basis for three-point functions involving such conserved tensors.

			The main result is a concrete independent basis for the tensor structures of three-point correlators of conserved operators in arbitrary 4D Lorentz representations, given in~\eqref{main} and subsequent formulae.
			We derived this formula by analysing supersymmetric generalisations and then arguing that the results apply directly to the non-supersymmetric case. 
			However, 
			the results themselves, and even a self-contained argument for them,  can be expressed directly in position space without requiring reference to supersymmetry or analytic superspace. Thus, for the reader's convenience, we present them now and leave the proofs and technical details of (S)CFT in analytic superspace for section \ref{sec:coset}. 
			
			We have included a \textsf{Mathematica} file in our submission with a function that evaluates it explicitly.
			In section \ref{subsec:examples}, we discuss some relevant examples and comparisons with the literature.
			Finally, we apply the same ideas to three-point functions with a non-conserved insertion in section \ref{sec:nonconserved}. Further checks can be found in appendix \ref{app:CF}.

			\subsection{Coincident limit}

			When all three insertions are conserved operators and the total left- and right-handed spins match, i.e. $\sum_i s_i = \sum\bar{s}_i$, we find  that a complete basis of tensor structures can be constructed from the following observed requirement: in all coincident limits $x_{ij}\to 0$, the structures must have the following leading order behaviour
			\begin{equation}\label{eq:OPElimit}
				\llangle J_{s_1,\bar{s}_1}(x_1) J_{s_2,\bar{s}_2}(x_2)J_{s_3,\bar{s}_3}(x_3)\rrangle 
				= \frac{1}{|x_{ij}|^{2(1+p)}} (x_{ij})_{(\alpha_1 (\dota_1} \dots (x_{ij})_{\alpha_p) \dota_p)} \times T(x_j,x_k)\Big(1+O(x_{ij}\Big)\,,
			\end{equation}
			where $p$ is constrained by the values of $s_i$ and $\bar{s}_i$ and
			$T$ is a tensor function of $x_j$ and $x_k \neq x_i,x_j$, which carries the remaining indices. The above asymptotic behaviour also applies to cases with $s=\bar{s}=0$, which correspond to dimension $2$ scalars. We consider scalars of different dimension and other general non-conserved insertions in section \ref{sec:nonconserved}.
			
			We will  derive this constraint by reducing from analytic superspace result in section \ref{sec:coset}, though it would be interesting to fully understand its origin from CFT arguments alone. 
			We observe that the expression multiplying $T$ in~\eqref{eq:OPElimit} solves the conservation constraint at points $i$ and $j$ by itself. This can be seen easily by observing that it takes the form of a two-point function of massless higher spin (twist 1) chiral fermions $\langle \psi_{\alpha_1\dots \alpha_p} \bar{\psi}_{\dota_1\dots \dota_p}\rangle$. The massless spinors satisfy the Dirac equation $$\partial^{\dotb\alpha_1}\psi_{\alpha_1\dots \alpha_p}=0$$   which then implies~\eqref{eq:constraint} at points 1 and 2.
			As we will see this single constraint~\eqref{eq:OPElimit}  uniquely determines all tensor structures of correlators of three arbitrary conserved tensors, as we will discuss in the following subsection.

			For future consideration, we note that an analogous statement can be made in arbitrary dimensions, namely:
			\begin{equation}\label{eq:generald}
				\partial_{i\nu_1} \left(\frac{1}{|x_{ij}|^{(d-2)(1+p)}} \Pi_{\text{ST}}\left(x_{ij}^{\nu_1} x_{ij}^{\nu_1} \dots x_{ij}^{\nu_{p}}\right)\right)=0,
			\end{equation}
			where $\Pi_{\mathrm{ST}}$ is a projector onto the symmetric traceless representation of $\SO(d-1,1)$. 
			Indeed, while in this paper we only find and prove bases of three-point functions in the four-dimensional case (also including non-conserved insertions where the above gets modified), it is natural to conjecture that  in the limit $x_{ij}\to 0$, the leading term in the three-point tensor structures of conserved tensors transforms in the symmetric traceless representation of $\SO(d-1,1)$ in arbitrary dimensions $d$. 
			The construction of a basis of such structures in arbitrary dimensions would then follow by analogous methods to the ones carried out in this paper.

			\subsection{General construction}\label{ssubsec:construction}
			
			The constraint~\eqref{eq:OPElimit} is  sufficient to reconstruct all possible linearly independent conformally invariant tensor structures for arbitrary spins. 
			Starting from an arbitrary linear combination of the overcomplete basis of tensor structures \eqref{eq:genstructure} for $\tau_i=2$ $(i=1,2,3)$, we impose the limits $x_{ij}\to 0$ and equate the leading term to the fully symmetrised tensor of \eqref{eq:OPElimit}. 
			This constrains the  coefficients of this linear combination, with the number of independent solutions being equal to the number of three-point structures after imposing conservation. 
			The independent solutions are shown to have a very simple formula which can be compactly written for arbitrary $s_i$ and $\bar{s}_i$ (subject to $\sum_i s_i -\bar{s}_i=0$). 
			In this section, we first outline the procedure for the simpler case where one insertion is a scalar and then show the general result (see~\eqref{main}). 
			
			\subsubsection{$\langle J_{s_1,\bar{s}_1}J_{s_2,\bar{s}_2} \phi^2\rangle$}

			This is a prototypical example of conservation constraints because they reduce the number of independent tensor structures from $\min(s_i,\bar{s}_i)+1$ (arising from~\eqref{noind3}) to just $1$, for all $s_1,\bar s_1,s_2, \bar s_2$.

			Let us first illustrate the construction for the first non-trivial example: two symmetric traceless spin-1 fields (currents), which were already studied in the 70s~\cite{Schreier:1971um}.
			Before imposing conservation, we have two tensor structures
			\begin{equation}\label{eq:JJphibefore}
				\langle J_{\alpha_1 \dota_1}(x_1) J_{\alpha_2 \dota_2 }(x_2) \phi^2(x_3)\rangle \ = \ \frac 1{x_{12}^2x_{13}^2x_{23}^2} \left( c_1 I_{12}{}_{\dota_1 \alpha_2} I_{21}{}_{\dota_2 \alpha_1} +c_2 Y_1{}_{\dota_1\alpha_1} Y_2{}_{\dota_2 \alpha_2} \right).
			\end{equation}
			The only relevant OPE limit with non-trivial tensors at leading order is $x_{12}\to 0$. The fact that only this limit is relevant is consistent with the fact that $\phi^2$ is not a conserved field, i.e. it is unconstrained. 
			The pole symmetrisation requirement \eqref{eq:OPElimit} on the limit of \eqref{eq:JJphibefore} requires that all indices appear symmetrised, so 
			\begin{equation}\label{JJphiasym}
				\langle J_{\alpha \dota}(x_1) J_{\beta \dotb }(x_2) \phi^2(x_3)\rangle \  = 
				\ \frac C{x_{12}^2x_{13}^4} \Big( (x_{12}^{-1})_{\dota_1\alpha_1} (x_{12}^{-1})_{\dota_2 \alpha_2} + (x_{12}^{-1})_{\dota_1 \alpha_2} (x_{12}^{-1})_{\dota_2 \alpha_1} \Big)
				\Big(1+O(x_{12})\Big).
			\end{equation}

			Now as $x_{12}\to 0$ we have that
			\begin{equation}\label{YIlim}
				V_{12}=I_{12}\to x_{21}^{-1}, \qquad V_{21}=-I_{21} \to x_{21}^{-1}, \qquad V_{11}=Y_1 \to x_{21}^{-1}, \qquad V_{22}=Y_2 \to x_{21}^{-1}.
			\end{equation}
			and plugging this into~\eqref{eq:JJphibefore} and comparing with~\eqref{JJphiasym} 
			implies that $c_1=-c_2$, giving a single tensor structure and hence OPE coefficient.
			
			For arbitrary $s_i,\bar{s}_i$ at points 1 and 2, the result  can be deduced similarly. We have the following basis of tensor structures before imposing conservation
			\begin{equation}\label{eq:s1s2phi}
				\langle J_{s_1,\bar s_1} J_{s_2,\bar s_2} \phi^2 \rangle = \frac1{x_{12}^2x_{13}^2x_{23}^2} \sum_i c_i (V_{11})^i  (V_{21})^{s_1-\bar s_1+i}(V_{11})^{\bar s_1-i}(V_{22})^{s_2-i}, 
				\qquad s_1+s_2=\bar s_1+\bar s_2.
			\end{equation}
			Then, letting $ J_{s_i,\bar{s}_i}=J_{(\alpha_{i,1}\dots\alpha_{i,s_i}),(\dota_{i,1}\dots\dota_{i,\bar s_i})}$, the asymptotic pole symmetrisation \eqref{eq:OPElimit} requires that
			\begin{align}\label{JJphiasym2}
				\langle J_{s_1,\bar s_1} J_{s_2,\bar s_2} \phi^2 \rangle
				\overset{x_{12}\to0}{\sim} C(x_{21}^{-1})^{s_1+s_2}_{(\alpha_{1,1}\dots\alpha_{1,s_1}\alpha_{2,1}\dots\alpha_{2,s_2})(\dota_{1,1}\dots\dota_{1,\bar s_1}\dota_{2,1}\dots\dota_{2,\bar s_2})}\ ,
			\end{align}
			as $x_{12}\to 0$.
			Imposing this limiting behaviour on~\eqref{eq:s1s2phi} using~\eqref{YIlim} completely fixes the functional form of the 3-point function. In fact, it is straightforward to directly uplift from this asymptotic behaviour~\eqref{JJphiasym2} to the full result away from the limit: simply replace the $x_{21}^{-1}$ terms according to their index structure as follows
			\begin{align}
				\begin{aligned}
					(x_{21}^{-1})_{\dota_1 \alpha_1} &\mapsto (V_{11})_{\dota_1 \alpha_1}, \qquad & \qquad (x_{21}^{-1})_{\dota_2 \alpha_2} &\mapsto (V_{22})_{\dota_2 \alpha_2},  \\
					(x_{21}^{-1})_{\dota_1 \alpha_2} &\mapsto (V_{12})_{\dota_1 \alpha_2},\qquad&   \qquad (x_{21}^{-1})_{\dota_2 \alpha_1} &\mapsto (V_{21})_{\dota_2 \alpha_1}    . 
				\end{aligned}
			\end{align}
			Thinking through the combinatorics this gives a specific combination of the terms in~\eqref{eq:s1s2phi} with a single unfixed coefficient:
			\begin{align}
				c_i=  \frac{C }{i!(\bar s_1{-}i)!(s_2{-}i)!(s_1{-}\bar s_1{+}i)!} 
				\, .
			\end{align}

			\subsubsection{$\langle J_{s_1,\bar{s}_1} J_{s_2,\bar{s}_2} J_{s_3,\bar{s}_3}\rangle$}\label{sec:generalresult}
			The result for the general case of three arbitrary spin conserved currents (with $\sum_i s_i=\sum_i \bar s_i$) can be deduced in a similar way. One insists that in all three separate limits $x_{ij}\rightarrow 0$ the limit looks like~\eqref{eq:OPElimit}. 
			Any given fixed structure~\eqref{eq:genstructure} then belongs  to three unique separate chains, according to its asymptotic behaviour in the three limits.  Then, each chain will look like~\eqref{eq:s1s2phi}, giving three sums.
			All in all, the result can be summarised in a remarkably simple analytic expression
			\begin{align}\label{main}
				\llangle J_{s_1,\bar{s}_1}J_{s_2,\bar{s}_2}J_{s_3,\bar{s}_3}\rrangle^{(a)} =\frac1{x_{12}^2x_{13}^2x_{23}^2} \sum_{c_1,c_2,c_3} \prod_{i,j=1}^3 \frac{V_{ij}^{b_{ij}}}{b_{ij}!}, \qquad \qquad \sum_i s_i=\sum_i \bar s_i,
			\end{align}
			where we recall the $V_{ij}$  defined in \eqref{Vs} and we define the exponents as
			\begin{align}\label{bsol}
				b_{ij}=\left(
				\begin{array}{ccc}
					a {-}c _2{-}c _3{+}s_1 & c _3 & \bar{s}_1{-}a {+}c _2{-}s_1 \\
					c _3{-}a  & \bar{s}_2{+}a{-}c _1{-}c _3 & c _1 \\
					c _2 & {-}\bar{s}_2{-}a {+}c _1{+}s_2 & {-}\bar{s}_1{+}a {-}c _1{-}c
					_2{+}s_1{+}s_3 \\
				\end{array}
				\right)_{ij}\ .
			\end{align}
			The sum in~\eqref{main} is over $c_i$, with $a$ held fixed. 
			The triple sum in $c_1,c_2,c_3$ is finite as it is constrained by $b_{ij}\geq 0$, but it can be formally left unrestricted since it will in any case be cut off by the terms $1/b_{ij}!$.\footnote{It can useful computationally to  cut them off by eg $c_1\leq s_3, c_2 \leq s_1, c_3 \leq s_2$. }
			The sum can be rephrased in a parametrisation independent way as the sum over the 
			set
			$S_{a}$ representing the set of positive integers $b_{ij}$ satisfying the following constraints
			\begin{align}\label{eq:Sa}
				S_a = \{b_{ij}\in \mathbb{Z}^{\geq 0}:\sum_{i=1}^3b_{ij}=s_j,\ \sum_{i=1}^3b_{ji}=\bar s_j,\ b_{12}-b_{21}=a\}\,.
			\end{align}
			We have attached a \textsf{Mathematica} file to the submission with the main formula above and several examples. In the next subsection, we show the most relevant ones.
			Note that the case $s_3=\bar{s}_3=0$ reproduces \eqref{eq:s1s2phi} (up to an overall normalisation), since only $a=0$ gives a non-zero result.

			To show that all structures generated in this manner are linearly independent, note that they are constructed from symmetrised products of the basic tensor invariants $V_{ij}$. For balanced tensor structures (satisfying \eqref{eq:balanced}), the only identity arises from the antisymmetric combination \eqref{idV} giving \eqref{eq:2} and the most general balanced higher-spin identity \eqref{eq:HSids}. Thus, our basis structures are orthogonal to this identity. Then, since they each have different powers of each $V_{ij}$ by construction and the only identity that can lower this power does not apply, they are linearly independent.

			Now we note that the sum in~\eqref{main} is  non-empty if and only if $a$ takes values between
			\begin{equation}\label{alims}
				\max(-s_1,-\bar s_2,\bar s_1{-}s_1{-}s_3)\leq a\leq \min(\bar s_1,s_2,\bar s_1{+}\bar s_3{-}s_1)\,.
			\end{equation}
			Therefore  the counting of tensor structures (different values of $a$ giving non-zero structures in \eqref{main}) can be deduced from~\eqref{alims} to be
			\begin{equation}\label{eq:counting}
				\dim\left( \langle J_{s_1,\bar{s}_1} J_{s_2,\bar{s}_2}J_{s_3,\bar{s}_3}\rangle\right) = \min_{i\neq j} (s_i+\bar{s}_i,s_i+s_j,\bar{s}_i+\bar{s}_j)+1,
			\end{equation}
			for all $s_i,\bar{s}_i$ such that $s_1+s_2+s_3=\bar{s}_1+\bar{s}_2+\bar{s}_3$.
			This is the general formula for the counting of structures after imposing conservation. Special cases of this result have appeared as conjectures in \cite{Costa:2011mg,Stanev:2012nq,Buchbinder:2023coi}. 
			So far we have only technically shown that this is a lower bound for the number of structures.
			With the coset space framework of section \ref{sec:coset}, we show that this method produces all possible tensor structures consistent with conservation.
			We also provide an alternative derivation of this counting formula in section \ref{sec:LRcounting} from Littlewood-Richardson coefficients, with more details in appendix \ref{app:LR}.
			
			For three symmetric traceless insertions ($s_i=\bar{s}_i$), we note that $a$ runs from $-\min(s_i)$ to $\min(s_i)$ in integer steps, giving $2\min(s_i)+1$ structures, as expected from the conjecture of \cite{Costa:2011mg}. Furthermore, these can be separated into parity even and odd as follows
			\begin{equation}\label{eq:oddVSeven}
				\begin{aligned}
					&\text{Even:} &\llangle J_{s_1} J_{s_2} J_{s_3}\rrangle^{(a)} +\llangle J_{s_1} J_{s_2} J_{s_3}\rrangle^{(-a)}, \\
					&\text{Odd:} &\llangle J_{s_1} J_{s_2} J_{s_3}\rrangle^{(a)} -\llangle J_{s_1} J_{s_2} J_{s_3}\rrangle^{(-a)}.
				\end{aligned}
			\end{equation}
			
			Finally it is straightforward to check  that the solution~\eqref{main} does indeed satisfy  the pole symmetrisation~\eqref{eq:OPElimit} we used to obtain it. Consider poles in $x_{12}^{-1}$. The poles  arise from $V_{11},V_{12},V_{21},V_{22}$ with  the other $V_{ij}$ are non singular when $X_{12} \rightarrow 0$. But we see from the parametrisation that the sum factorises over these terms as
			\begin{align}
				\sum_{\gamma_1,\gamma_2} \left(\sum_{\gamma_3} \frac{V_{11}^{\beta_{11}}V_{12}^{\beta_{12}}V_{21}^{\beta_{21}}V_{22}^{\beta_{22}} }{\beta_{11}!\beta_{12}!\beta_{21}!\beta_{22}!} \right) \frac{V_{13}^{\beta_{13}}V_{31}^{\beta_{31}}V_{23}^{\beta_{23}}V_{32}^{\beta_{32}}V_{33}^{\beta_{33}}}{\beta_{13}!\beta_{31}!\beta_{23}!\beta_{32}!\beta_{33}!}\ .
			\end{align}
			The term in brackets then gives the combination in \eqref{eq:s1s2phi} and the remaining terms are independent of $c_3$. Similar arguments apply to the other $x_{23}^{-1}$ and  $x_{13}^{-1}$ poles.\footnote{In fact there are potential signs to consider in the $X_{13}^{-1}$ case. Indeed, while $V_{11},V_{22}\to x_{13}^-1$, we have  that $V_{13}=I_{13} ,V_{31}=-I_{31} \to  -x_{13}^{-1}$. However one can see from~\eqref{bsol} that   $b_{13}-b_{31}$ is fixed within a tensor structure and so all terms in it will have the same number of $I_{13}$s minus $I_{31}$s. Thus 
				flipping the sign of $I_{13}$ and $I_{31}$  can only at most give an overall minus sign change in the definition of the tensor structure, not affecting its conservation.} 
			
			The construction of tensor structures for correlators with non-zero imbalance $\sum_i s_i-\bar{s}_i$ does not directly follow from the OPE limit of \eqref{eq:OPElimit}. However, as we discuss in section \ref{sec:coset}, the general analytic superspace method used to derive the asymptotic behaviour \eqref{eq:OPElimit} and thus the main result \eqref{main} also applies to such unbalanced correlators.

			\subsection{Examples}\label{subsec:examples}

			\subsubsection{Currents $\langle JJJ\rangle$} 
			By \eqref{eq:genstructure}, we have the following list of (overcomplete) tensor structures before imposing conservation
			\begin{equation}\;
				\langle J_{\alpha \dota}(x_1) J_{\beta \dotb }(x_2) J^{\gamma \dotc}(x_3)\rangle \ = \ P_2 \sum_{i=1}^{6} c_i  \     \begin{pmatrix}
					I_{12\,\dota \bb}\, I_{23\,\dotb \gamma}\, I_{31\,\dotc \alpha}\\
					I_{21\,\dotb \alpha}\, I_{32\,\dotc \bb}\, I_{13\,\dota \gamma}\\
					Y_{1\,\dota \alpha}\, I_{23\,\dotb \gamma}\, I_{32\,\dotc \beta}\\
					Y_{2\,\dotb \beta}\, I_{13\,\dota \gamma}\, I_{32\,\dotc \beta}\\
					Y_{3\,\dotc \gamma}\, I_{12\,\dota \beta}\, I_{21\,\dotb \alpha}\\
					Y_{1\,\dota \alpha}\, Y_{2\,\dotb \beta}\, Y_{3\,\dotc \gamma}
				\end{pmatrix}.
			\end{equation}
			Unlike in the previous case, the three OPE limits yield non-trivial tensor poles which have to be symmetrised. This produces constraints on the above $c_i$ coefficients analogous to the differential conservation constraints. All in all, the three possible combinations consistent with \eqref{eq:OPElimit} are
			\begin{equation}\label{eq:JJJ}
				\begin{aligned}
					a=1&: & \llangle J^{\alpha \dota}_1 J^{\beta \dotb }_2 J^{\gamma \dotc}_3\rrangle^{(1)}&= I_{12}^{\dota \bb} I^{\dotb \gamma}_{23} I_{31}^{\dotc \alpha},\\
					a=0&: &\llangle J^{\alpha \dota}_1 J^{\beta \dotb }_2 J^{\gamma \dotc}_3\rrangle^{(0)}&= Y^{\dota \alpha }_1 Y^{\dotb \beta}_2 Y^{\dotc \gamma}_3 - \left(Y^{\dota \alpha}_1 I^{\dotb \gamma}_{23}I_{32}^{\dotc \beta} + Y^{\dotb \beta}_2 I^{\dota \gamma}_{13}I_{31}^{\dotc \alpha}+Y^{\dotc \gamma}_3 I^{\dota \beta}_{12}I_{21}^{\dotb \alpha}\right),\\
					a=-1&: &\llangle J^{\alpha \dota}_1 J^{\beta \dotb }_2 J^{\gamma \dotc}_3\rrangle^{(-1)}&=-I_{21}^{\dotb \alpha} I^{\dotc \bb}_{32} I_{13}^{\dota \gamma},
				\end{aligned}
			\end{equation}
			where the cases $a=\pm 1$ correspond to a single pole $x_{ij}^{-1}$ in all $x_{ij}\to0$ limits, and the case $a=0$ yields a double (symmetrised) pole. Note the minus sign in the $a=-1$ term is not necessary but comes from evaluating \eqref{main} and is important for splitting into even and odd components as in \eqref{eq:oddVSeven}.
			To match with the basis in the CFTs4D package \cite{Cuomo:2017wme}, we note the identity~\eqref{eq:2}. We discuss such checks in more detail in appendix \ref{app:CF}.

			\subsubsection{Stress tensors $\langle TTT\rangle$}
			For higher values of spin, the results can be presented in an index-free way as discussed in \ref{sec:corr}. 
			The following is the explicit evaluation of \eqref{main} for three stress tensors:
			\begin{equation}\label{eq:TTT}
				\begin{aligned}
					a=\pm2&: & \llangle TTT\rrangle^{(\pm2)}&= \frac{1}{8}\left\{ \begin{aligned}&I_{12}^2 I_{23}^2 I_{31}^2\\
						&I_{21}^2 I_{32}^2 I_{13}^2
					\end{aligned} \right. \ , \\
					a=\pm1&: & \llangle TTT\rrangle^{(\pm1)}&= \frac{1}{2}\left\{ \begin{aligned}&2 Y_1 Y_2 Y_3 I_{12}I_{23}I_{31}-\left(Y_2 I_{31}^2 I_{23} I_{13} I_{12} + \text{ cyclic } \right)\\
						&-2 Y_1 Y_2 Y_3 I_{21}I_{32}I_{13}+\left(Y_2 I_{13}^2 I_{32} I_{31} I_{21} + \text{ cyclic } \right)
					\end{aligned} \right.\ , \\
					a=0&: & \llangle TTT\rrangle^{(0)}&= \frac18 Y_1^2 Y_2^2Y_3^2 - I_{12}I_{21} I_{23}I_{32} I_{31} I_{13}-\frac12 \left(Y_3^2 Y_1 Y_2 I_{12} I_{21} + \text{ cyclic }\right) \\
					& & & \qquad + \frac18 \left(Y_1^2 I_{23}^2 I_{32}^2+ \text{ cyclic } \right) + \left( Y_1 Y_2 I_{13}I_{31}I_{23}I_{32} + \text{ cyclic } \right)\ .
				\end{aligned}
			\end{equation}
			Taking the different OPE limits we see that \eqref{eq:OPElimit} holds. Indeed, note that the terms $a=\pm2$, $a=\pm 1$, and $a=0$ correspond to double, triple and quadruple symmetrised poles, respectively. Using equation \eqref{eq:oddVSeven}, the above basis can be written in terms of purely parity even and odd structures.
			
			By going to the conformal frame of the CFTs4D package \cite{Cuomo:2017wme} (to avoid manually using identities between basic covariants such as \eqref{eq:2}), one can easily check that our results are consistent and provide an alternative basis for $\langle TTT \rangle$. We show this in appendix \ref{app:CF}. 
			\subsubsection{Beyond}
			The main formula \eqref{main} works for any correlator of three conserved tensors in 4D.
			We now demonstrate its power by evaluating it for a high spin mixed symmetry correlator:
			\begin{equation}
				\langle J_{2,5} J_{10,4}J_{3,6}\rangle.
			\end{equation}
			This operation is evaluated instantly since the formula is analytic.
			We have attached a \textsf{Mathematica} file with this and several other examples to our submission, including the full implementation of \eqref{main} for the reader's convenience.
			Below, we list the first 3 structures out of the expected 6 from \eqref{alims} and \eqref{eq:counting} ($a=0,\dots,5$):
			\begin{align}
				\llangle J_{2,5} J_{10,4}J_{3,6}\rrangle^{(0)}&=- Y_{1}^{2} Y_{2}^{4} I_{13}^{3} I_{32}^{6}
				+ 8\, Y_{1} Y_{2}^{3} I_{12} I_{13}^{3} I_{21} I_{32}^{6}
				- 6\, Y_{2}^{2} I_{12}^{2} I_{13}^{3} I_{21}^{2} I_{32}^{6},\\ \notag
				\llangle J_{2,5} J_{10,4}J_{3,6}\rrangle^{(1)}&=Y_{1}^{2} Y_{2}^{4} Y_{3} I_{12} I_{13}^{2} I_{32}^{5}
				+ 72\, Y_{1} Y_{2}^{3} Y_{3} I_{12}^{2} I_{13}^{2} I_{21} I_{32}^{5}
				- 36\, Y_{2}^{2} Y_{3} I_{12}^{3} I_{13}^{2} I_{21}^{2} I_{32}^{5}\\
				& \qquad \notag
				+ 12\, Y_{1} Y_{2}^{4} I_{12} I_{13}^{3} I_{31} I_{32}^{5}
				- 24\, Y_{2}^{3} I_{12}^{2} I_{13}^{3} I_{21} I_{31} I_{32}^{5}
				+ 12\, Y_{1}^{2} Y_{2}^{3} I_{12} I_{13}^{2} I_{23} I_{32}^{6} \\\notag
				& \qquad 
				- 36\, Y_{1} Y_{2}^{2} I_{12}^{2} I_{13}^{2} I_{21} I_{23} I_{32}^{6}
				+ 12\, Y_{2} I_{12}^{3} I_{13}^{2} I_{21}^{2} I_{23} I_{32}^{6}, \\\notag
				\llangle J_{2,5} J_{10,4}J_{3,6}\rrangle^{(2)}&=-45\, Y_{1}^{2} Y_{2}^{4} Y_{3}^{2} I_{12}^{2} I_{13} I_{32}^{4}
				+120\, Y_{1} Y_{2}^{3} Y_{3}^{2} I_{12}^{3} I_{13} I_{21} I_{32}^{4}
				-45\, Y_{2}^{2} Y_{3}^{2} I_{12}^{4} I_{13} I_{21}^{2} I_{32}^{4}
				\\\notag
				& \qquad 
				+90\, Y_{1} Y_{2}^{4} Y_{3} I_{12}^{2} I_{13}^{2} I_{31} I_{32}^{4}
				-120\, Y_{2}^{3} Y_{3} I_{12}^{3} I_{13}^{2} I_{21} I_{31} I_{32}^{4}
				-15\, Y_{2}^{4} I_{12}^{2} I_{13}^{3} I_{31}^{2} I_{32}^{4}
				\\\notag
				& \qquad 
				+72\, Y_{1}^{2} Y_{2}^{3} Y_{3} I_{12}^{2} I_{13} I_{23} I_{32}^{5}
				-144\, Y_{1} Y_{2}^{2} Y_{3} I_{12}^{3} I_{13} I_{21} I_{23} I_{32}^{5}
				+36\, Y_{2} Y_{3} I_{12}^{4} I_{13} I_{21}^{2} I_{23} I_{32}^{5}\\\notag
				& \qquad 
				-72\, Y_{1} Y_{2}^{3} I_{12}^{2} I_{13}^{2} I_{23} I_{31} I_{32}^{5}
				+72\, Y_{2}^{2} I_{12}^{3} I_{13}^{2} I_{21} I_{23} I_{31} I_{32}^{5}
				-18\, Y_{1}^{2} Y_{2}^{2} I_{12}^{2} I_{13} I_{23}^{2} I_{32}^{6}\\\notag
				& \qquad 
				+24\, Y_{1} Y_{2} I_{12}^{3} I_{13} I_{21} I_{23}^{2} I_{32}^{6}
				-3\, I_{12}^{4} I_{13} I_{21}^{2} I_{23}^{2} I_{32}^{6}.
			\end{align}
			
		\subsection{One non-conserved operator}\label{sec:nonconserved}
			We now consider (balanced) three-point functions of two conserved fields and a third non-conserved field. 
			Constructing a basis of tensor structures for such correlators is slightly different to the three-conserved-insertions case, though they follow from a common argument in analytic superspace (see section \ref{sec:coset}). 
			
			The first step is
			to view the non-conserved operator in a basis that diagonalises the free theory OPE of two conserved currents, that is, identifying the operator appearing in the free $J\times J$ OPE, rather than writing it generically in terms of dimension and spin.
            
			For example, the free $J\times J$ OPE contains a $\Delta=6$ scalar $$:J^\mu J_{\mu}:\,,$$
            rather than  $\phi^6$. 
            Using  the fact that antisymmetrisation of 4D Weyl spinor indices corresponds to index contractions in 4D, in our formalism we write this as
			\begin{equation}
				J^\mu J_\mu \leftrightarrow J_{[\alpha[\dota} J_{\beta]\dotb]}\equiv\cO^4_{[\alpha \beta] [\dota \dotb]},
			\end{equation}
        where the superscript denotes the number of fundamental ($\tau=1$) scalar fields used to construct such an operator. 
        Then we solve the Ward identities  according to this
        free theory perspective according to the rules outlined in section~\ref{sec:4D}, namely soak up indices with $I_{ij}$ or $Y_i$ (with the indices antisymmetrised at point 3). So for example
			\begin{equation}\label{eq:nonconserved1}
				\begin{aligned}
					\langle J_{\alpha_1 \dota_1} J_{\alpha_2 \dota_2} \cO^{6}\rangle &\equiv \langle J_{\alpha_1 \dota_1} J_{\alpha_2 \dota_2} \cO^4_{[\alpha_3 \beta_3] [\dota_3 \dotb_3]}\rangle \ni \{I_{13}I_{23}I_{31}I_{32},Y_1Y_3I_{23}I_{32},Y_2Y_3I_{13}I_{31}\}.
				\end{aligned}
			\end{equation}


			To make this more precise, in a three-point function with two arbitrary conserved currents $J_{s_1,\bar{s}_1}$, $ J_{s_2,\bar{s}_2}$ with a third non-conserved operator $\cO^{\tau}_{s,\bar{s}}$ we view the third operator equivalently as follows
			\begin{equation}\label{otoo}
				\cO^{\tau}_{s,\bar{s}} \rightarrow \cO^4_{\ula_L, \ula_R} 
			\end{equation}
            where 
                       \begin{align}\label{eq:4D}
\notag
                   \ula_L&=[\tfrac12(\tau{-}4){+}s,\tfrac12(\tau{-}4)] & \ula_R&=[\tfrac12( \tau{-}4){+}\bar s,\tfrac12( \tau{-}4)] \\[0.2cm]
                      &=\begin{tikzpicture}[baseline=(align)]
					\def\topwidth{2}   %
					\def\botwidth{1.0}   %
					\def\rowheight{0.3}  %
					\draw[line width=0.6pt] (0,\rowheight) rectangle (\topwidth,2*\rowheight);
					\draw[line width=0.6pt] (0,0) rectangle (\botwidth,\rowheight);
					\draw[<->] (1.03,-0.15) -- (\topwidth,-0.15)
					node[midway, below] {$s$};
					\node (align) at (0,0.2) {};
				\end{tikzpicture}\ ,&  &=\begin{tikzpicture}[baseline=(align)]
					\def\topwidth{2.3}   %
					\def\botwidth{1}   %
					\def\rowheight{0.3}  %
					\draw[line width=0.6pt] (0,\rowheight) rectangle (\topwidth,2*\rowheight);
					\draw[line width=0.6pt] (0,0) rectangle (\botwidth,\rowheight);
					\draw[<->] (1.03,-0.15) -- (\topwidth,-0.15)
					node[midway, below] {$\bar{s}$};
					\node (align) at (0,0.2) {};
				 \end{tikzpicture} \ .
           \end{align}
These are two-row Young diagrams filled with undotted and dotted Weyl spinor indices, respectively which  are now allowed to be antisymmetrised, as encoded by the second row.
            We see that  the length of each row depends on the spins of the conserved fields as well as on twist and spin of the non-conserved field itself.
            
            The above reformulation is a convenient way of organising the contributions of all non-conserved operators appearing in the $J \times J$ OPE.
            But it naively  looks  like we are making things more complicated, since we are introducing more indices and thus structures have to be constructed from more covariants. 
            However, the key advantage is that, as we shall show in the next section using analytic superspace techniques, conservation conditions at points 1 and 2 imply the following remarkably simple key constraint. When the operators are viewed as~\eqref{otoo}:  
			\begin{align}\label{eq:nonconservedconstraint}
				\text{\em Tensor structures must be completely independent of  $x_{12}^{-1}$}\,.   
			\end{align}
In particular the tensor structures can only be constructed from the following basic covariants
			$$I_{31},\  I_{13},\ I_{23},\ I_{32},\ Y_3,$$
			thus significantly reducing the allowed space of tensor structures.
			so we have that 
			\begin{equation}\label{eq:noncons}
				\llangle J_{s_1,\bar{s}_1}J_{s_2,\bar{s}_2} \cO^4_{\ula_L,\ula_R}\rrangle = \frac{1}{(x_{23}^2)^2(x_{31}^2)^2} \Pi_{\ula_L,\ula_R}(x_3) \prod_{ij} V_{ij}^{b_{ij}}, \qquad b_{ij}= \begin{pmatrix}
					0 & 0 & b_{13} \\
					0 & 0 & b_{23} \\
					b_{31} & b_{32} & b_{33}
				\end{pmatrix},
			\end{equation}
			where $\Pi_{\ula_L,\ula_R}(x_3)$ projects all spinor indices at point 3 according to the Young tableaux $\ula_L$, $\ula_R$ of the non-conserved insertion (in the free-field basis) and where
			\begin{equation}
				\begin{aligned}
					b_{13}+b_{23}+b_{33} &= |\ula_L|, \\
					b_{31}+b_{32}+b_{33} &= |\ula_R| .
				\end{aligned}
			\end{equation}
We note that there will in general be multiple different projectors $\Pi_{\ula_L,\ula_R}(x_3)$ for a given choice of Young diagrams $\ula_L,\ula_R$. Each of them should be thought of as giving a different linearly independent tensor structure. However, this number is always smaller than the naive $\dim \ula_L \dim \ula_R$ due to the symmetries already imposed at points 1 and 2, the constraint that $V_{33}$ must transform in the same representation on both left and right indices, as well as additional weaker pole symmetrisation constraints in $x_{13}$ and $x_{23}$. We discuss some general examples below.
            
            The non-conserved insertion in \eqref{eq:noncons} can be converted to the generic basis in terms of total dimension and symmetrised Weyl spinor indices using \eqref{eq:4D}.
			Furthermore, because indices at the third insertion are often antisymmetrised, thus producing determinants, the above tensor structures can be written in terms of purely symmetrised indices. The most convenient way to derive identities between \eqref{eq:noncons} and the more standard form for tensor structures (not in the basis of free fields) is to pick a conformal frame. See appendix \ref{app:CF} for details. We now show some examples. 
			
			Firstly consider the motivating example \eqref{eq:nonconserved1}. Notice that only one of the possible structures listed on the RHS is independent of $x_{12}$ and so using \eqref{eq:noncons} we find that only this unique structure survives
			\begin{equation}\label{eq:JJphi6}
				\llangle J_{\alpha_1 \dota_1} J_{\alpha_2 \dota_2} \cO^4_{[\alpha_3 \beta_3] [\dota_3 \dotb_3]}\rrangle= \frac{1}{(x_{23}^2)^2(x_{31}^2)^2}(I_{31})_{[\dota_3 \alpha_1} (I_{32})_{\dotb_3] \alpha_2} (I_{13})_{\dota_1 [\alpha_3} (I_{23})_{\dota_2 \beta_3]}\,.
			\end{equation}
			This should be compared to  the standard approach where there are two tensor structures whose relative weight needs to be determined by the conservation condition.
			Indeed one can show (most easily  by picking a conformal frame as described in appendix~\ref{app:CF}), that~\eqref{eq:JJphi6} can be rewritten in terms of these two tensor structures as follows
			\begin{equation}\label{eq:JJphi6v2}
				\llangle J J \phi^6\rrangle = \frac{x_{12}^2}{(x_{13}^2)^3 (x_{23}^2)^3}\left( I_{12} I_{21} + Y_1 Y_2\right),
			\end{equation}
			agreeing with known results. We show this explicitly in appendix \ref{app:CF}.

			For correlators of two symmetric traceless conserved currents and a scalar of arbitrary dimension the Ward identities are solved by
			\begin{equation}
				\llangle J_s J_s \cO^{2(2+s+\delta)} \rrangle \overset{\eqref{eq:4D}}{=}\llangle J_s J_s \cO^{4}_{[(s+\delta)^2],[(s+\delta)^2]} \rrangle =  \frac{1}{(x_{23}^2)^2(x_{31}^2)^2} \Pi_{[(s+\delta)^2]}(x_3) \left(I^s_{31}I^s_{32}Y_3^{2\delta} I^s_{13} I^s_{23} \right),
			\end{equation}
			where $\Pi_{[c^2]}(x_3)$ denotes a projector to representation $[c^2]=[c,c]$ of the indices at point 3, both undotted and dotted separately (implements index contraction). Equivalently in 4D all indices at point 3 are contracted away with $\epsilon^{\alpha \beta}$s in all possible ways, leaving the indices at points 1 and 2 completely symmetrised. 
            When $\delta=0$ there is only one such inequivalent projector (the result is given by products of the spin 1 structure~\eqref{eq:JJphi6}, symmetrised over its dotted and undotted indices at points 1 and 2). For higher dimension operators at point three however, i.e. with $\delta>0$, there is more than one inequivalent projector. Further constraints occur however from considering the leading terms when $x_{13}\rightarrow 0$ and $x_{23} \rightarrow 0$ and a similar constraint  to that of~\eqref{eq:OPElimit} must be imposed in these limits. In this context, rather than constraining the relevant indices to be completely symmetrised, here the constraint is that they must lie in two-row Young diagrams. In fact, the leading term in the limit would vanish if this were not the case, since we can't have three-row Young diagrams in $\SL(2)$. Therefore,  this constraint  can be rephrased as saying that the leading term must not vanish in the limits $x_{13},x_{23}\rightarrow 0 $. Just as for the conserved current case~\eqref{eq:OPElimit}, this constraint originates in considerations of analytic superspace as detailed in the next section.

			If the third insertion is also symmetric traceless then the result is very similar
              \begin{equation}\label{eq:example}
                \begin{split}
   				\llangle J_s J_s \cO^{4+2s+2\delta}_{s'} \rrangle^{(i)} &\overset{\eqref{eq:4D}}{\equiv} \llangle J_s J_s \cO^{4}_{[s+\delta+s',s+\delta]} \rrangle^{(i)}  \\&= \frac{1}{(x_{23}^2)^2(x_{31}^2)^2} \Pi^{(i)}_{[s+\delta+s',s+\delta]}(x_3) \left(I_{31}^sI_{32}^sY_3^{2\delta+s'} I_{13}^s I_{23}^s\right).
                \end{split}
			\end{equation}
            As in the $s'=0$ case there are a number of inequivalent projectors. Imposing the asymptotic condition in the limits $x_{13},x_{23} \rightarrow 0$ reduces the number of independent terms down further to 
			\begin{equation}\label{eq:JJOscount}
				\dim (\langle J_{s} J_{s}\cO_{s'}\rangle) =  \min(s,s')+1,
			\end{equation}
            independently of $\delta$.
            This number can be derived from combinatorial arguments involving Littlewood-Richardson coefficients as we show in the next section and in appendix \ref{app:LR}.

			All tensor structures counted by \eqref{eq:JJOscount} are linearly independent because the projectors $\Pi(x_3)$ in \eqref{eq:example} are orthogonal by definition and they symmetrise according to two-row Young diagrams, i.e. there is no antisymmetrisation of 3 Weyl spinor indices which would produce identities as discussed in section \ref{sec:identities}.

          As outlined in appendix \ref{app:CF}, all structures can be written in a more common basis using known identities, which can be most easily derived by picking a frame.
          Furthermore, while our derivation uses combinatorics thus requiring integer values of $\delta$ (contributing to the dimension), all results can be analytically continued to non-integer values, yielding the correct result. This is possible by introducing quasi-tensors already used in a different context in \cite{Heslop:2001gp,Heslop:2001zm}.
          We leave details of this procedure for future work.

			\section{Coset space methods for conformal correlators}\label{sec:coset}
			All of the results described in the previous section have been obtained by reducing 4D CFT from a general theory with $\SU(m,m|2n)$ symmetry. In particular, $(m,n)$ analytic superspace~\cite{Galperin:1984av,Howe:1995md,Heslop:2001gp,Heslop:2003xu,Doobary:2015gia,Hansen:2025lig} 
			provides a unified description independent of $m,n$, in which only the size of the coordinate matrix changes, not the form of the tensor structures. We begin this section with a brief review of this formalism (see \cite{Hansen:2025lig} for a more detailed treatment using similar notation).

			Then, we highlight the key aspect of this unified formalism. Namely that as shown in \cite{Heslop:2001zm}, all representations with $n\neq 0$ are always irreducible, meaning shortening (e.g. conservation) conditions are automatically satisfied at the unitarity bound. This is due to the strong constraints imposed by compactness of the `$\mathrm{R}$-symmetry' subgroup $\SU(2n)$. Indeed, analyticity in the coordinates associated with this subgroup (internal coordinates $y$) automatically solves all shortening conditions.
			At the level of correlation functions, imposing analyticity on the space of possible tensor structures is then equivalent to imposing the relevant conservation constraints. Since the solution is independent of $m,n$ other than through the coordinate matrix, setting $n=0$ then solves the constraints in the non-supersymmetric 4D CFT case ($m=2$) by construction.
			This unified setup can then be used to derive general statements about non-supersymmetric CFT kinematics from supersymmetric arguments.
			
			Furthermore, by setting $m=0$, thus considering a compact $\SU(2n)$ theory where fields are given by finite-dimensional representations, we can map the problem of counting conformal tensor structures to Littlewood-Richardson coefficients.

			\subsection{Unified coset construction}\label{sec:cosetspace}
			Four-dimensional conformal field theory on Minkowski space is a special case of large class of coset spaces known as analytic superspaces \cite{Howe:1995md} which have been used to study superconformal field theories primarily due to the unconstrained nature of the superfields \cite{Heslop:2001zm,Heslop:2003xu}, a feature not present in other superspace constructions. More recently, they have been used to derive (super)conformal blocks \cite{Doobary:2015gia,Aprile:2021pwd,Heslop:2022xgp,Aprile:2025nta,Hansen:2025lig}. 
			
			We construct $(m,n)$ analytic superspace as a coset space of the complexified superconformal group $\SU(m,m|2n)\overset{\mathbb{C}}{\to}\SL(2m|2n)$ of the form
			\begin{equation}\label{eq:coset}
				\SL(m|2n|m)/ H\,, 
			\end{equation}
			where $H$ is the (parabolic) subgroup of lower block triangular matrices with the choice of columns $m|2n|m$. We label coset spaces by the pair of integers $(m,n)$. The 4D conformal case we consider in this paper then corresponds to $(2,0)$.  We note that the $(m,n)$ coset space \eqref{eq:coset} should be intuitively understood as the general class of homogeneous spaces on which $\SL(2m|2n)$ acts analogously to how the 4D conformal group $\SU(2,2)$ acts on Minkowski space.
			
			Coordinates on \eqref{eq:coset} are given by square $(m|2n|m)$ matrices with an equivalence relation due to the coset. One can use this to show that 
			\begin{equation}
				\SL(m|2n|m)/ H \cong \mathrm{Gr}(m|n,2m|2n),
			\end{equation}
			i.e. a (super-)Grassmannian of $(m|n)$-planes in $\mathbb{C}^{2m|2n}$. Homogeneous coordinates on the Grassmannian are denoted $X^{\cA}_{\ \fa}$. It is also useful to consider the  orthogonal plane $\bX^{\cA \dfa}$ so we have
			\begin{equation}\label{eq:XbX}
				X^{\cA}_{\ \fa}, \ \bX^{\cA \dfa}, \qquad \qquad  X^{\cA}_{\ \fa}\bX^{\cA \dfa} =0,
			\end{equation}
			where $\cA$ is a $\SL(m|2n|m)$ fundamental (i.e. a superconformal index) and $\fa,\dfa$ are $\SL(m|n)$ indices. The analytic superspace coordinate 
			\begin{equation}\label{eq:analyticX}
				\mathrm{x}^{\fa \dfa} =\begin{pmatrix}
					x^{\alpha \dota} & \rho^{\alpha a'} \\
					\bar{\rho}^{a \dota} & y^{a a'}
				\end{pmatrix},
			\end{equation}
			can be recovered from an open chart as follows
			\begin{equation}\label{eq:superspaceframe}
				X^{\cA}_{\ \fa} \sim \begin{pmatrix}
					1_{m|n} & \mathrm{x}^{\fa \dfa}
				\end{pmatrix}, \qquad \qquad  \bX^{\cA\dfa} \sim \begin{pmatrix}
					-\mathrm{x}^{\fa \dfa}\\
					1_{m|n}
				\end{pmatrix}.
			\end{equation}
			
			\subsection{Fields}\label{sec:cosetfields}
			Fields on the coset space \eqref{eq:coset} are equivariant maps 
			\begin{equation}
				\cO:\SL(2m|2n)\rightarrow V \ \text{such that }\  \cO(g\cdot h) = \cO(g) \cdot R(h), \qquad h\in H, \, g\in \SL(2m|2n),
			\end{equation}
			where $V$ is a representation space of $H$. Thus, they  transform in a finite-dimensional representation of $H$, which is block lower triangular and so splits into the block diagonal subgroup known as the Levi subgroup:
			\begin{equation}\label{eq:Levi}
				L:=\SL(m|n) \times \SL(m|n) \times \mathbb{C}^*.
			\end{equation}
			This simply is a generalisation of the 4D conformal representations where fields transform with Weyl spinor indices for $\SL(2)\times \SL(2)$ and a charge encoding the scaling dimension. Indeed, a $(2,0)$ coset space field is labelled by a charge $\gamma$ and two $\SL(2)$ Young diagrams $\ula_L, \ula_R$ which map to the usual labels for a conformal multiplet $s,\bar{s},\Delta$ using \eqref{eq:4D}
			
			In the general $(m,n)$ case, $\SL(m|n)$ Young diagrams are subdiagrams of a thick hook with $m$ rows of infinite length and $n$ columns of infinite height
			\begin{equation}
				\begin{tikzpicture}[x=14pt, y=14pt,scale=1,baseline=(current bounding box.center)]
					\draw[line width=0.6pt] (0,0) -- (8,0);   %
					\draw[line width=0.6pt] (0,0) -- (0,-5);  %
					
					\draw[line width=0.6pt] (2,-2) -- (8,-2);   %
					\draw[line width=0.6pt] (2,-2) -- (2,-5);   %
					\draw[<->] (-0.25,0) -- (-0.25,-2)
					node[midway, left] {$m$};
					
					\draw[<->] (0,0.25) -- (2,0.25)
					node[midway, above] {$n$};
					\node at (7.5,-1) {$\dots$};   %
					\node at (1,-4.5) {$\vdots$}; 
				\end{tikzpicture}\ .
			\end{equation}
			
			All in all a field on the coset space will be labelled by the representation of its highest weight state under the Levi subgroup ($\SL(m|n)$ Young diagrams and a charge $\gamma$) as follows
			\begin{equation}\label{eq:mnfield}
				\mathcal{V}^{2m|2n}([\ula_L,\ula_R]_{\gamma}) \equiv \cO^\gamma_{\ula_L,\ula_R},
			\end{equation}
			where $\mathcal{V}$ denotes the set of descendants of the highest weight state (equivalently, the induction from $[\ula_L,\ula_R]_{\gamma}$ into a full $\SL(2m|2n)$ representation). This will be infinite when $m\neq 0$ (i.e. non-compact conformal side is turned on) and finite when $m=0$.
			
			What each Young diagram means depends on what values of $m,n$ one is interested in (e.g. $m=2$ corresponds to 4D $\cN=2n$). See \cite{Hansen:2025lig} for the full translation between coset space fields and 4D (super)conformal representations. However, in practice the values of these integers does not have to be fixed until the end of the calculations, allowing for $m,n$-independent statements.  This is best seen when considering invariants on \eqref{eq:coset}, i.e. correlation functions.
			
			\subsection{$\SL(2m|2n)$ Ward identities}\label{sec:cosetWIs}
			To construct multi-point invariant functions, we note that $\SL(2m|2n)$ acts linearly on the $(m,n)$ coset space \eqref{eq:coset}. It thus suffices to contract the $\cA,\mathcal{B}$ indices. The resulting functions will be covariant under the Levi subgroup \eqref{eq:Levi} but superconformally invariant. 
			\paragraph{Expectation values:} As is known in (S)CFT, one-point functions vanish unless the operator is the identity (trivial Levi representation). This is satisfied on the coset space for all $m,n$ due to the orthogonality of $X$ and $\bX$ \eqref{eq:XbX}.
			So any invariant function of Grassmannian variables $X, \bX$ must be a constant, i.e. transform in the trivial representation of \eqref{eq:Levi}, as expected.
			\paragraph{Two points:} Out of two points, we can now take the following non-zero invariant
			\begin{equation}
				X_{ij}^{\fa \dfa}:= (X_i)^{\fa}_{\ \cA} (\bX_j)^{\cA \dfa} \sim (\mathbf{x}_i - \mathbf{x}_j)^{\fa \dfa}:= \mathbf{x}_{ij}^{\fa \dfa},
			\end{equation}
			with inverse and superdeterminant (Berezinian)
			\begin{equation}\label{eq:twopointinvariants}
				(X_{ij}^{\fa \dfa})^{-1} := (\hX_{ji})_{\dfa \fa} \sim (\mathrm{x}^{-1}_{ij})_{\dfa \fa}, \qquad \qquad g_{ij}:= \mathrm{sdet} (\hX_{ij}) \sim \frac{\det y_{ji}}{\det x_{ji}}+ O(\rho, \bar{\rho}).
			\end{equation}
			We note that $\hX_{ij}$ is a direct $(m,n)$ generalisation of the inversion tensor $I_{ij}$ in both position and embedding space approaches to 4D CFT. A two-point function of general coset fields $\cO^\gamma_{\ula_L,\ula_R}$ is thus given by the unique tensor structure
			\begin{equation}\label{eq:2pt}
				\langle \cO^\gamma_{\ula_L, \ula_R} \cO^\gamma_{\ula_R, \ula_L} \rangle = C g_{12}^\gamma \left[\hX_{12}^{\otimes |\ula_L|}\right]_{\ula_L}\left[\hX_{21}^{\otimes|\ula_R|}\right]_{\ula_R}.
			\end{equation}
			
			\paragraph{Three points:} 
			To cover the whole space of three-point invariants in analytic superspace it suffices to introduce the following tensor
			\begin{equation}\label{eq:Y}
				(Y_{i,jk})_{\dfa \fa} := (\hX_{ij})_{\dfa \fb}X_{jk}^{\fb \dfb} (\hX_{ki})_{\dfb \fa} \sim (\mathbf{x}_{ji}^{-1} + \mathbf{x}_{ik}^{-1})_{\dfa \fa}.
			\end{equation}
			We have that $Y_{i,jk}=-Y_{i,kj}$ and thus we identify $Y_i$ with the positive permutation of $i,j,k$.
			This is a direct $(m,n)$ generalisation of our 4D structures $V_{ij}$ \eqref{Vs}. 
			Three-point tensor structures for correlators of $\SL(2m|2n)$ fields are given by direct analogy with the 4D expression \eqref{eq:genstructure}:
			\begin{equation}\label{eq:3pt}
				\llangle \cO_1 \cO_2 \cO_3 \rrangle = g_{12}^{\gamma_{12,3}} g_{23}^{\gamma_{23,1}} g_{31}^{\gamma_{31,2}} \prod_{i,j,k=1}^{3}Y_k^{b_{k}} \hX_{ij}^{b_{ij}},
			\end{equation}
			where $\gamma_{ij,k}=(\gamma_i+\gamma_j-\gamma_k)/2$ and the open indices should be symmetrised locally according to the representations at each insertion.
			See \cite{Heslop:2003xu,Hansen:2025lig} for more details on the general construction, including that of nilpotent invariants. 
			
			To construct tensor structures for unbalanced correlators ($B=\sum_i (s_i {-}\bar s_i)/2\neq 0$ in \eqref{eq:4Dbalance}) in 4D, we must  dress the balanced tensor structures with appropriate unbalanced covariants
			\begin{equation}\label{eq:KKb}
				(K_{ij,k})^{\fa}_{\ \fb} := X_{ik}^{\fa \dfa}(\hX_{kj})_{\dfa \fb}, \quad (B>0),\qquad \qquad (\bar{K}_{ij,k})_{\dfa}^{\ \dfb} := (\hX_{ik})_{\dfa \fa} X_{kj}^{\fa \dfb}, \quad (B<0),
			\end{equation}
			as in the 4D case discussed in section \ref{sec:corr}. Interestingly, for generic $m,n \neq 0$, there is no epsilon tensor that raises or lowers indices, and upper indices in such cases correspond to non-unitary superconformal multiplets \cite{Heslop:2001zm}.
			
			The key feature of the $(m,n)$ coset space is that, when written in terms of these basic covariants, tensor structures are universal, i.e. they take the same form independently of $m$ and $n$. Indeed, $m,n$ only control the sizes of the matrices and the degree of antisymmetrisation that is allowed. 
			Fixing values of $m,n$ produces identities between combinations of covariants, making the basis \eqref{eq:3pt} overdetermined, as discussed for 4D ($m=2,n=0$) in section \ref{sec:corr}. 
			This redundancy can be fixed by writing the structures in terms of analytic superspace variables \eqref{eq:superspaceframe} and picking a frame, analogously to the use of a conformal frame in ordinary CFT. 
			Interestingly, conservation constraints yield the same number of linearly independent structures independently of $m$ and $n$, as we will see next.
			
			\subsection{Conservation constraints \& automatic irreducibility}
			The $\SU(m,m|2n)$ unitarity bound that generalises that of conserved tensors in 4D CFT corresponds to $\gamma=2$ and one-row Young diagrams $\ula_L=[s],\ \ula_R=[\bar{s}]$, written as
			\begin{equation}\label{eq:conservedsuperfield}
				\cO^{\gamma=2}_{[s],[\bar{s}]}=\cO^{2}_{(\fa_1 \dots \fa_s)(\dfa_1 \dots \dfa_s)},
			\end{equation}
			where the indices are supersymmetrised: symmetrised Weyl spinors $x_{\alpha \dota}$ but antisymmetrised internal indices $y_{a a'}$ as follows
			\begin{align}\label{indexsym}
				(\fa \fb)=\tfrac12\left\{ \begin{array}{ll}
					\alpha \beta+\beta \alpha  \qquad& \fa=\alpha, \fb=\beta \\
					\alpha b+b\alpha &\fa=\alpha, \fb=b\\
					a \beta +\beta a&\fa=a, \fb=\beta\\
					ab-ba&\fa=a, \fb=b
				\end{array}     \right.\ ,
			\end{align}
			so we view the indices $a,b$ as Grassmann odd.
			Using the translation \eqref{eq:4D} for 4D fields $(m=2)$ we find that $\gamma=\Delta-(s+\bar{s})/2$ is the twist $\tau$, whose unitarity bound is $\tau\geq2$, as expected. 
			
			The generalised conservation constraint for the field \eqref{eq:conservedsuperfield} in terms of the analytic superspace coordinate is
			\begin{equation}\label{eq:mnconservation}
				\partial_\mathrm{x}{}_{\fa \dfa}\cO^2_{(\fa_1 \dots \fa_s)(\dfa_1 \dots \dfa_{\bar{s}})}=  \partial_\mathrm{x} {}_{(\fa (\dfa}\cO^2_{\fa_1 \dots \fa_s)\dfa_1 \dots \dfa_{\bar{s}})},
			\end{equation}
			i.e. all the indices must be supersymmetrised. This is just a rephrasing of the vanishing antisymmetrisation, which in 4D ($m=2$) is a contraction of indices, giving the usual conservation constraint \eqref{eq:constraint}.
			In terms of the Grassmannian coordinates $X,\bX$ it is given by
			\begin{equation}\label{eq:mnconservation2}
				(\partial_{\bX})_{\dfa \cA} (\partial_{X})^{\cA}_{\ \fa} \cO^2_{(\fa_1 \dots \fa_s)(\dfa_1 \dots \dfa_{\bar{s}})}=(\partial_{\bX})_{(\dfa \cA} (\partial_{X})^{\cA}_{\ (\fa} \cO^2_{\fa_1 \dots \fa_s)\dfa_1 \dots \dfa_{\bar{s}})},
			\end{equation}
			which can directly be applied to any set of tensor structures constructed in the previous section. However, as shown in \cite{Heslop:2001zm}, this condition is automatically satisfied in analytic superspace ($n \neq 0$). The reason is that the null states associated with the highest-weight state of the constraint always lie inside the Levi subgroup, and are thus automatically accounted for since we directly take irreducible representations of this subgroup.
			It is crucial for this that the internal space ($n\neq0$) is compact. Thus fields on this compact subgroup are automatically irreducible. In terms of the analytic superspace coordinate $\mathrm{x}_{\fa \dfa}$, they are polynomials in the internal variable $y_{a a'}$, with homogeneity carried by $\gamma$.\footnote{Note this also ensures consistency across charts.}

			So once we have found a basis of analytic three-point tensor structures for arbitrary $m,n$ in a way that depends on $m,n$ only through the size of the vector spaces carried by the (super)-indices, then we have found a solution of the Ward identities, which also  satisfies the conservation condition~\eqref{eq:mnconservation}. Then, in the $(2,0)$ case this must also satisfy the 4D CFT Ward identities and conservation conditions!

			\subsection{Conserved correlators}\label{sec:conservedcorrcoset}
			
			The key idea is that all correlation functions must be polynomial in the internal part of the Grassmannian coordinates  when $n\neq 0$ 
			and then  conservation conditions are automatically satisfied at each point.

			This is a non-trivial constraint since  as all the basic covariants ($\hX_{ij}$, $Y_i$) depend on inverse coordinates $y_{ij}^{-1}$. The resulting singularity must be cancelled by the determinant prefactor appearing in the $N$-point function. This prefactor is a product of powers of superdeterminants (Berezinians) of the superspace coordinate $\mathrm{x}_{ij}$ \eqref{eq:twopointinvariants}, determined by the charge $\gamma$ at each insertion. Thus, their numerator contains appropriate powers of $\det(y_{ij})$ which can cancel (tensor) poles in $y_{ij}^{-1}$ under certain conditions.
			
			Naively, one would expect to need as many powers of $\det(y_{ij})$ as there are tensors $y_{ij}^{-1}$ to cancel each pole individually. However, if different tensors appear with antisymmetrised indices, this softens the pole. 
			More explicitly, consider an ordinary (non super) $n\times n$ matrix $M$ (which for us will be the matrices $y_{ij}$). Then, an antisymmetrised product of $p$ copies of its inverse $M^{-1}$ only has a single pole in $\det M$, rather than a $p$th order pole. Indeed 
			\begin{align}
				(M^{-1})^{k_1}_{[l_1}\dots (M^{-1})^{k_p}_{l_p]} =
				\frac1{\det M} \epsilon_{i_1\dots i_n}M^{i_{p+1}}_{j_{p+1}}\dots M^{i_n}_{j_n}\epsilon^{j_1\dots  j_n}\frac1{(n-p)!},
			\end{align}
			so only a single $\det M$ pole appears on the right  
			rather than the naively expected $\det^p M$.\footnote{This formula follows directly from $\det M \epsilon^{i_1\dots i_n}=M^{i_1}_{j_1}\dots M^{i_n}_{j_n}\epsilon^{j_1\dots j_n}$.}
			
			\subsubsection{Two points}
			The general two point function is given by \eqref{eq:2pt} with prefactor
			\begin{equation}
				g_{12}^\gamma \sim \frac{\det(y_{12})^\gamma}{\det(x_{12})^\gamma}.
			\end{equation}
			For the generalised $(m,n)$ conserved operators $\cO^2_{[s],[\bar{s}]}$ \eqref{eq:conservedsuperfield}, the two point function takes the form
			\begin{equation}
				\langle \cO^2_{[s],[\bar{s}]}\cO^2_{[\bar{s}],[s]}\rangle = C g_{12}^2 (\hX_{12}^{\otimes\bar{ s}})_{\text{sym}}(\hX_{21}^{\otimes s})_{\text{sym}}.
			\end{equation}
			Now recall that symmetrisation of superindices $\fa, \dfa$ corresponds to antisymmetrisation of internal indices $a, a'$ (see~\eqref{indexsym}), thus the tensor structure carries two poles of $\det y_{12}$, but the prefactor cancels them, giving a fully analytic structure in the internal variables. One can explicitly check that the above two-point function satisfies the conservation constraint \eqref{eq:mnconservation2} as expected.

			\subsubsection{Three points}
			The above process applied to three point functions produces tensor structures that satisfy the general conservation constraint \eqref{eq:mnconservation2}. The $(m,n)$ three-point function \eqref{eq:3pt} contains the following prefactor
			\begin{equation}\label{eq:prefactor}
				P_{\gamma_1,\gamma_2,\gamma_3}:= g_{12}^{\gamma_{12,3}} g_{23}^{\gamma_{23,1}} g_{31}^{\gamma_{31,2}} \sim \frac{\det(y_{12})^{\gamma_{12,3}} \det(y_{23})^{\gamma_{23,1}}\det(y_{31})^{\gamma_{31,2}} }{\det(x_{12})^{\gamma_{12,3}} \det(x_{23})^{\gamma_{23,1}}\det(x_{31})^{\gamma_{31,2}}}.
			\end{equation}
			
			For three conserved insertions $(\gamma_i=2)$, all $\gamma_{ij,k}=1$, that is
			\begin{equation}
				P_{2,2,2}=g_{12}g_{23} g_{31} \sim \frac{\det(y_{12}) \det(y_{23})\det(y_{31}) }{\det(x_{12}) \det(x_{23})\det(x_{31})}.
			\end{equation}
			There is only one determinant of each $y_{ij}$ and thus to obtain consistent tensor structures, all tensor dependence on each of the $y_{ij}$ must be antisymmetrised (supersymmetrised in $\fa,\dfa)$. This gives the $(m,n)$ superspace origin of the asymptotic behaviour \eqref{eq:OPElimit} we presented in the results section, namely
			\begin{equation}\label{eq:mnOPElimit}
				\llangle \cO^2_{[s_1],[\bar{s}_1]}(\mathrm{x}_1) \cO^2_{[s_2],[\bar{s}_2]}(\mathrm{x}_2)\cO^2_{[s_3],[\bar{s}_3]}(\mathrm{x}_3)\rrangle \overset{\mathrm{x}_{ij}\to0}{\sim} \mathrm{sdet}(\mathrm{x}^{-1}_{ij} ) (\mathrm{x}^{-1}_{ij})_{(\fa_1 (\dfa_1} \dots (\mathrm{x}^{-1}_{ij})_{\fa_p) \dfa_p)}.
			\end{equation}
			Setting $m=0$ the right hand side gives a polynomial in the $y_{ij}$
			\begin{equation}
				\det(y_{ij}) (y_{ij}^{-1})_{[a_1 [a'_1 }\dots (y_{ij}^{-1})_{a_p] a'_p] } = \frac{1}{(n-p)!} (y_{ij})^{a_{p+1} a'_{p+1}} \dots (y_{ij})^{a_{n} a'_{n}} \epsilon_{a'_1 \dots a'_n} \epsilon_{a_1 \dots a_n},
			\end{equation}
			as expected.
			Constructing tensor structures then becomes a pole cancellation exercise, which can be easily fixed by noting the $y^{-1}$ dependence of each of the basic covariants $\hX_{ij}$ \eqref{eq:twopointinvariants} and $Y_i$ \eqref{eq:Y}.
			
			The general solution to the pole cancellation exercise for three conserved unitary superfields \eqref{eq:conservedsuperfield} with arbitrary one-row Young diagrams $\ula_L=[s_i]$, $\ula_R=[\bar{s}_i]$ is
			essentially identical to our main result for conserved tensor structures in a 4D CFT ~\eqref{main}, with all elements interpreted in the general $m,n$ context. So  it can be written in terms of  generalised covariants $V_{ij}$ defined by~\eqref{Vs}  but in terms of  supercovariants $I_{ij}{}_{\dota \alpha}\mapsto \hX_{ij}{}_{\dfa \fa}$ \eqref{eq:twopointinvariants} and $Y_{i}{}_{\dota \alpha} \mapsto Y_{i}{}_{\dfa \fa}$ \eqref{eq:Y}:
			\begin{align}\label{maincoset}
				\llangle \cO^{\gamma=2}_{[s_1],[\bar{s}_1]}\cO^{\gamma=2}_{[s_2],[\bar{s}_2]}\cO^{\gamma=2}_{[s_3],[\bar{s}_3]}\rrangle^{(a)} =g_{12} g_{23} g_{31}\sum_{S_a} \prod_{i,j=1}^3 \frac{V_{ij}^{b_{ij}}}{b_{ij}!},
			\end{align}
			where $S_a$ is given in \eqref{eq:Sa} (equivalently \eqref{bsol}).
			Remarkably, all structures are linearly independent for all $m,n$ unlike non-conserved structures which satisfy additional $m,n$-dependent identities (arising from the $\SL(m|n)$ epsilon tensor, e.g. antisymmetrising $m+1$ indices in $\SL(m)$).
			The structures automatically satisfy the general $m,n$ conservation constraint \eqref{eq:mnconservation}, i.e. they also satisfy the 4D constraint ($m=2,n=0$) by construction on the unified coset space. Mapping back to 4D thus gives the main formula \eqref{main}. 
			
			The same formula can instead be interpreted as 4D $\cN\!=\!2$  ($m\!=\!2,n\!=\!1$) or $\cN\!=\!4$ ($m\!=\!2,n\!=\!2$) theories by translating the labels appropriately. This gives the analogous formula for the space of superconformally invariant three-point tensor structure of higher-spin supercurrents (twist two operators).
			The same pole cancellation exercise can be carried out for other representations arising from mixed symmetry diagrams arising in the $\SL(2|1)$ and $\SL(2|2)$ portions of the Levi subgroup \eqref{eq:Levi}, e.g. quarter-BPS representations in $\cN=4$ which have totally antisymmetric Young diagrams $\ula_L,\ula_R$.
			
			The OPE of two conserved superfields $\cO^2_{[s],[\bar{s}]}$ in general contains several non-conserved operators with $\gamma=4$ and arbitrary two-row Young diagrams
			\begin{equation}\label{eq:gamma4}
				\cO^4_{[\lambda_1, \lambda_2], [\mu_1, \mu_2]} \in \cO^2_{[s_1],[\bar{s}_1]}\times\cO^2_{[s_2],[\bar{s}_2]}.
			\end{equation}
			Using the conversion \eqref{eq:4D}, these fields map to the expected operators in the OPE of two 4D (higher-spin) currents in the basis that diagonalises the free field, i.e. with explicit antisymmetrisation of indices indicating index contraction, as explained in section \ref{sec:nonconserved}. 
			The scalar prefactor \eqref{eq:prefactor} of three-point functions of such fields is now given by
			\begin{equation}\label{eq:224prefactor}
				P_{2,2,4}=g_{12}g_{23} g_{31} \sim \frac{ \det(y_{23})^2\det(y_{31})^2 }{ \det(x_{23})^2\det(x_{31})^2},
			\end{equation}
			where the non-conserved field is at point 3. To be consistent with analyticity in $y_{ij}$ the tensor structures must not have any explicit $y_{12}$ dependence when written as coset fields \eqref{eq:gamma4}. This is the origin of the constraint \eqref{eq:nonconservedconstraint} on 4D correlators with a non-conserved third insertion $\llangle JJ \cO \rrangle$, leading to the result in equation \eqref{eq:noncons}. Furthermore, any $y_{23}^{-1}$ and $y_{13}^{-1}$ dependence must still be cancelled away by the determinants in \eqref{eq:224prefactor}. Because there are two powers of each determinant, the dependence on $y_{23}^{-1}$ and $y_{13}^{-1}$ must be symmetrised according to a two-column Young diagram. This further reduces the number of structures as outlined in section \ref{sec:nonconserved}.
			
			Here we have considered correlators of coset fields for unitary superconformal ($n>0$) representations, which have lower indices only \cite{Heslop:2001zm}. These must then satisfy
			\begin{equation}
				\sum_{i=1}^3 |\ula_L^i|=\sum_{i=1}^3 |\ula_R^i|,
			\end{equation}
			where $\ula_L^i$ and  $\ula_R^i$ are Young diagrams for the Levi subgroup \eqref{eq:Levi} representation at each insertion. In 4D, these translate to correlators satisfying $\sum_i s_i = \sum \bar{s}_i$ \eqref{eq:balanced}, i.e. correlators built out of $I_{ij}$ and $Y_i$ only, not $K_{ij}$ and $\bar{K}_{ij}$ \eqref{eq:KK}.
			The construction of such `unbalanced' tensor structures follows from the same pole cancellation exercise albeit on non-unitary coset space representations. We leave a detailed study of these cases for future work.

			\subsection{Counting conformal invariants via Littlewood-Richardson rule}\label{sec:LRcounting}
			Because the number of solutions to the conservation constraints is independent of $m$ and $n$, we can now reduce to the {\em purely compact}  representation-theoretic arguments from the $(m,n)=(0,n)$ coset space to quickly count the number of conformally invariant structures in 4D $(m,n)=(2,0)$. 
			This reduces CFT kinematic questions to finite-dimensional counting problems in $\SL(2n)$! The catch is that you need $n$ to be large enough to contain the representations you are interested in.
			
			Indeed, the counting of these structures is mapped to Littlewood-Richardson coefficients of $\SL(2n)$.
			The calculations can be carried out analytically for all spins, which leads to an alternative proof of the general counting formula \eqref{eq:counting}. More details, including a review of the representation theory tools being used, can be found in appendix \ref{app:LR}.
			
			For $m=0$, the conserved superfield \eqref{eq:conservedsuperfield}, when ruduce dot this purely compact space  becomes a field where the indices are {\em antisymmetrised}
			\begin{equation}
				\cO^{2}_{\begin{tikzpicture}[scale=0.25, baseline=(current bounding box.center)]
						\draw[line width=0.5pt] (0.5,0) rectangle (4,1);
						\node[right] at (2.9,0.55) {$\scriptstyle s$};
					\end{tikzpicture},\,\begin{tikzpicture}[scale=0.25, baseline=(current bounding box.center)]
						\draw[line width=0.5pt] (1.2,0) rectangle (4,1);
						\node[right] at (2.9,0.55) {$\scriptstyle \bar{s}$};
				\end{tikzpicture}} \overset{m=0}{\longrightarrow} \cO^{2}_{[a_1\dots a_s][a'_1\dots a'_{\bar{s}}]}.
			\end{equation}
			Since the real form of $\SL(|2n)$ we want is the compact group $\SU(2n)$, the above is a finite-dimensional representation, which can be more simply written as a single Young diagram. The precise form of the $\SL(2n)$ diagram is obtained from the induced representation $$\SL(|n)\times \SL(|n)\times\mathbb{C}^* \to \SL(|2n),$$ which gives the following identity between $\SL(2n)$ representations
			\begin{equation}\label{eq:conservedYD}
				\cO^{2}_{[a_1\dots a_s][a'_1\dots a'_{\bar{s}}]}\ =\ \begin{tikzpicture}[x=14pt, y=18pt,scale=1.2, baseline=(align)]
					\node (align) at (0.25,-1.7) {};
					\draw [ line width=0.6pt ] (0,0) rectangle (0.5,-2);
					\draw [ line width=0.6pt ] (0,-2) rectangle (0.5,-2.9);
					\draw [ line width=0.6pt ] (0.5,0) rectangle (1,-1);
					\node at (0.25,-0.82) {$\vdots$};
					\draw [line width=0.6pt,<->] (-0.25,0) -- (-0.25,-1.95)node[midway,left]{$n$};]
					\draw [line width=0.6pt,<->] (-0.25,-2.05) -- (-0.25,-2.9)node[midway,left]{$\bar{s}$};
					\draw [line width=0.6pt,<->] (1.25,-1) -- (1.25,-2)node[midway,right]{$s$};
					\draw [line width=0.6pt,dashed] (-0.5,-2) -- (1.7,-2);
				\end{tikzpicture} \ := \uLa_{s_1\bar{s}_1} \ .
			\end{equation}
			In full generality, a $m,n$ coset (analytic superspace) field maps to a finite-dimensional $\SL(|2n)$ rep under $m=0$ as follows
			\begin{equation}\label{eq:SL2nYD}
				\cO^\gamma_{\ula_L, \ula_R} \ \overset{m=0}{\longrightarrow} \ \begin{tikzpicture}[scale=0.85, baseline=(align)]
					\draw[line width=0.6pt] (0,0) -- (0,-4) ;
					\draw[line width=0.6pt] (1,-4) -- (1,0);
					\draw [dashed, line width=0.6pt] (-0.5,-2)--(4.7, -2);
					\draw[line width=0.6pt] (4.4,0) -- (4.4,-0.6) -- (4,-0.6) -- (4,-1.2) -- (3.6,-1.2) -- (3.6,-1.6) -- (3.2,-1.6) -- (3.2,-2) --(1.8,-2) -- (1.8,-2.9) -- (1.4,-2.9) -- (1.4,-3.6) -- (1,-3.6);
					\draw[dotted,line width=0.8pt] (0,0) -- (0,0.5);
					\draw[dotted,line width=0.8pt] (1,0) -- (1,0.5);
					\draw[dotted,line width=0.8pt] (4.4,0) -- (4.4,0.5);
					\draw[dotted,line width=0.8pt] (0,-4) -- (0,-4.5);
					\draw[dotted,line width=0.8pt] (1,-4) -- (1,-4.5);
					\draw[line width=0.6pt] (1, -2) -- (1.8,-2);
					\node at (1.4, -2.5) {$\ula_R^T$};
					\node at (4, -1.5) {$\ula_L^T$};
					\draw [dashed, line width=0.6pt] (4.4,-2)--(4.4, -0.6);
					\draw [line width=0.6pt,->] (-0.3,0) -- (-0.3,-1.98)node[pos=0.4,left]{$n$};
					\draw [line width=0.8pt,dotted,<-] (-0.3,0.7) -- (-0.3,0);
					\draw [line width=0.6pt,->] (-0.3,-4) -- (-0.3,-2.02)node[pos=0.4,left]{$n$};
					\draw [line width=0.8pt,dotted,<-] (-0.3,-4.7) -- (-0.3,-4);
					\draw [line width=0.6pt,<->] (1,0.7) -- (4.4,0.7)node[midway,above]{$\gamma$};
					\draw [line width=0.6pt,<->] (1,-4.7) -- (0,-4.7)node[midway,below]{$c$};
					\node (align) at (0,-2) {};
				\end{tikzpicture} \ ,
			\end{equation}
			where $c$ is an arbitrary positive integer, which does not change the $\SL(2n)$ representation since the determinant representation is trivial.

			The OPE of two $m=0$ coset fields is nothing but a tensor product of the above Young diagrams, which is given by the Littlewood-Richardson (LR) rule \cite{Fulton_reps}
			\begin{equation}
				\uLa_1 \otimes \uLa_2= \bigoplus_{\uLa} \uLa^{\oplus C^{\uLa}_{\uLa_1,\uLa_2}},
			\end{equation}
			where $C^{\uLa}_{\uLa_1,\uLa_2}$ are Littlewood-Richardson coefficients, which count the multiplicity of $\uLa$ in $\uLa_1\otimes \uLa_2$. This number then precisely counts the number of independent tensor structures in the three-point function $\langle \uLa_1 \otimes \uLa_2 \otimes \uLa\rangle$, i.e. the number of possible ways to construct the trivial representation out of $\uLa_1,\uLa_2,\uLa$. Note the following must hold for the LR coefficient to be non-zero
			\begin{equation}\label{eq:requirementLR}
				|\uLa|=|\uLa_1|+|\uLa_2|,
			\end{equation}
			which in practice requires non-zero values of $c$ in \eqref{eq:SL2nYD}.
			
			For example, consider a three point function of the simplest conserved coset fields
			\begin{equation}
				\langle \cO^2_{\Box,\Box} \cO^2_{\Box,\Box} \cO^2_{\Box,\Box}\rangle,
			\end{equation}
			which in 4D $(m=2)$ corresponds to $\langle JJJ\rangle$. In the compact space $(m=0)$, the field is given by the Young diagram \eqref{eq:conservedYD} with $s=\bar{s}=1$. In order for \eqref{eq:requirementLR} to hold, we must pad one of the diagrams with $c=1$ as in \eqref{eq:SL2nYD}. The corresponding LR coefficient is
			\begin{equation}
				C^{[1^{2n},\uLa_{1,1}]}_{\uLa_{1,1},\uLa_{1,1}} = \left\{ \begin{aligned}
					&n,  & n&<3,\\
					&3, & n&\geq 3. \\
				\end{aligned}\right. 
			\end{equation}
			In general, LR coefficients are known to be stable as $n$ grows, meaning they reach a constant value at finite $n$. This value of $n$ is called the stability threshold. The stable value of the LR coefficient reproduces the counting for coset spaces with non-zero $m$ (i.e. gives the number of solutions to the conservation constraint, equivalently, the pole cancellation exercise outlined in the previous section). Indeed, a correlation function of three currents is well-known to have 3 conformally invariant tensor structures, 2 parity-even and 1 parity-odd. Similarly, the counting for $\langle TTT\rangle$ is given by the following LR coefficient
			\begin{equation}
				C^{[1^{2n},\uLa_{2,2}]}_{\uLa_{2,2},\uLa_{2,2}} = \left\{ \begin{aligned}
					&n-1,  & n&<6,\\
					&5, & n&\geq 6, \\
				\end{aligned}\right. 
			\end{equation}
			giving the expected counting for correlators of stress-tensors in 4D (3 even, 2 odd).

			Note that for `large' $\SL(2n)$ Young diagrams (both rows and columns), LR coefficients can be hard to compute in general (though powerful software exists \cite{buch_lrcalc}). However, the finite-dimensional representations associated with conserved tensors \eqref{eq:conservedsuperfield} are restricted to two columns, and are thus far from being large. This makes it possible to derive LR coefficients for arbitrary values $s_i$ and $\bar{s}_i$. We have left such counting proofs to appendix \ref{app:LR}.
			The main result is
			\begin{equation}\label{eq:counting2}
				C^{\uLa_{s_3,\bar{s}_3}}_{\uLa_{s_1,\bar{s}_1}\uLa_{s_2,\bar{s}_2}} = \min_{i\neq j}(s_i+\bar{s}_i, s_i+s_j, \bar{s}_i+\bar{s}_j)+1, \qquad \forall \ n\geq \sum_{i=1}^{3} \frac{s_i+\bar{s}_i}{2},
			\end{equation}
			assuming the diagrams have been padded appropriately to ensure \eqref{eq:requirementLR} holds. This stable value then gives the counting of structures for
			\begin{equation}
				\langle J_{s_1,\bar{s}_1} J_{s_2,\bar{s}_2}J_{s_3,\bar{s}_3}\rangle,
			\end{equation}
			agreeing with \eqref{eq:counting}.
			
			For a non-conserved third insertion, padding is not necessary since they correspond to diagrams with 4 columns, i.e. $\gamma=4$ in \eqref{eq:SL2nYD}. The stability threshold is a lot simpler, it corresponds to $n$ such that the isotropy Young diagrams $\ula_L$, $\ula_R$ of the non-conserved insertion exist. All in all, the following is the counting of a correlator of two conserved fields and a third non-conserved one:
			\begin{equation}
				\dim \left(\langle J_{s_1,\bar{s}_1}J_{s_2, \bar{s}_2} \cO^4_{\ula_L,\ula_R}\rangle\right) = C^{\cO^4_{\ula_L,\ula_R}}_{\uLa_{s_1,\bar{s}_1} \uLa_{s_2,\bar{s}_2}}, \qquad \forall \ n>\max\left({\ula_L^{(1)}, \ula_R^{(1)}}\right),
			\end{equation}
			where $\ula^{(1)}$ denotes the (length of) the first row of $\ula$, and $\cO^{4}_{\ula_L,\ula_R}$ should be interpreted as \eqref{eq:SL2nYD} for computing the LR coefficients.
			
			Finally, we note that all the above counting formulae can be reinterpreted for $(m,n)=(2,1)$ and $(2,2)$ to yield the counting of 4D $\cN=2$ and $\cN=4$ superconformally invariant three-point tensor structures. Similar exercises can be performed for other isotropy Young diagrams $\ula_L$, $\ula_R$ which are relevant to such SCFTs, e.g. $\cN=4$ quarter-BPS supermultiplets which correspond to two columns of equal length $\ula_{L,R}=[1^{h}]$.
			
			\section{Conclusion \& outlook}
			\label{conclusion}

			In this work, we set up a general procedure for obtaining the basis of (conserved or non-conserved) three-point tensor structures in 4D CFT.
			The main result is the derivation of a concrete and compact formula~\eqref{main} for a complete basis of three-point tensor structures for balanced correlation functions (satisfying \eqref{eq:balanced}). 
			This followed directly from imposing the simple constraint~\eqref{eq:OPElimit} on the leading singular terms in each of the coincident limits $x_i \rightarrow x_j$. 
			We further showed that similar constraints~\eqref{eq:nonconservedconstraint}  apply to correlators involving a single non-conserved insertion, leading to a non-standard basis of tensor structures~\eqref{eq:noncons}. 
			It would be interesting to derive analytic formulae analogous to~\eqref{main} for such non-conserved cases as well as for unbalanced correlators.

			The constraints on coincident limits were shown to have a common origin in the lift to $\SU(m,m|2n)$ analytic superspace. Indeed, in this framework, the (lifted) conservation constraints are satisfied by imposing analyticity in the internal $\SU(2n)$ variables, and thus three-point structures are constructed via a systematic pole cancellation exercise. 
			Finally, we developed a novel method for counting three-point conformal invariants by mapping them to the well-understood Littlewood-Richardson coefficients. This and the construction \eqref{main} provide two different proofs (and generalisations) of various conjectures on the counting of conserved invariants \cite{Costa:2011mg, Buchbinder:2023coi}.
			Due to the unified formalism, all our results and methods can be directly reinterpreted to yield 4D $\cN=2$ and $\cN=4$ superconformally invariant three-point tensor structures of non-half-BPS operators, which are analogous to spinning operators in CFT, with half-BPS being analogous to scalars.

			There are many possible avenues for future work which we summarise now.

We emphasise that our method singles out a unique canonical basis, whose elements differ by their behaviour in coincident limits.
It would be interesting to compare this with other bases defined in the literature, most notably the helicity or amplitude basis of~\cite{Caron-Huot:2021kjy,Lee:2023qqx}, which relates the structures to scattering amplitudes in a QFT of one dimension higher. In particular, one could explore the relationship between our basis and amplitudes in 5D, for example using the spinor-helicity formalism developed in~\cite{Chiodaroli:2022ssi,Pokraka:2024fao}.

			A natural future direction is the generalisation to CFT in other dimensions. Here it is natural to  postulate that there exists a basis of tensor structures of conserved fields in arbitrary dimensions that is symmetric traceless in the various OPE limits, due to \eqref{eq:generald}. Then one could  construct tensor structures from this constraint as in section \ref{sec:results}. One would also of course want to prove the constraint for which the idea  would be to set up an analogous coset space to \eqref{eq:coset} but applied to conformal groups in other dimensions. For example, a unified coset space for certain 3D and 6D SCFT can be constructed as
			\begin{equation}
				\mathrm{OSp}(4n|2m)/P,
			\end{equation}
			which is formulated analogously to our 4D coset space. Such coset spaces were recently used to derive half-BPS (super)conformal blocks in \cite{Aprile:2021pwd}. 
			The idea would then be completely analogous to that developed in the present work: pole cancellation in the internal coordinates gives the uplift of the symmetric traceless OPE limit and solves the conservation constraint.

			The study of (3D) CFT has recently been applied to cosmology and inflationary physics, since certain inflationary models are given by an approximate de Sitter spacetime \cite{Strominger:2001pn, McFadden:2009fg,Maldacena:2011nz,McFadden:2011kk,Arkani-Hamed:2015bza}. To connect with observational data, expressions for CFT correlators must be derived in momentum space \cite{Bzowski:2013sza,Bzowski:2017poo,Bzowski:2018fql}. The question of whether there exists an analogous formula to \eqref{main} in terms of momentum variables is left for future work.

			In this paper, we have only considered unitary representations.
			However, non-unitary fields, particularly the so-called \emph{partially} conserved tensors have very recently been studied in the context of de Sitter holography in \cite{Baumann:2017jvh,Baumann:2025tkm} (also see earlier work \cite{Dolan:2001ih, Skvortsov:2006at}). These are traceless symmetric conformal fields with twist $\tau\leq1$ satisfying weaker differential constraints, namely
			\begin{equation}
				\partial_{\mu_1} \dots \partial_{\mu_p} \cO^{\mu_1 \dots \mu_p \dots \mu_s} = 0.
			\end{equation}
			In the bulk, they are dual to partially massless fields with some leftover gauge symmetry that reduces the number of possible polarisations.
			These representations are non-unitary in an AdS bulk but consistent in dS, hence the renewed interest.
			We expect both our OPE asymptotics and analytic superspace constructions to apply to this kind of representations, since partial conservation can be straightforwardly lifted to arbitrary $m$ and $n$ within our unified coset framework. It would thus be interesting to construct a formula analogous to \eqref{main} for three-point functions involving partially conserved tensors and the stress tensor. Such a result could help determine the consistent couplings of partially massless fields to gravity at large scales in our universe.
			
			Another obvious direction is to superconformal field theory. 
			As mentioned above, by reinterpreting the main result \eqref{maincoset} for $n=1$ and $n=2$, it yields tensor structures for 
			4D $\cN=2$ and $\cN=4$ superconformally invariant twist-two three-point structures. 
			However, it would be interesting to  generalise to other representations.
			The counting of tensor structures can be straightforwardly performed for any representation from Littlewood-Richardson coefficients as outlined in section \ref{sec:LRcounting}. For example this  gives that a three-point correlator of quarter-BPS superconformal multiplets $\cO_{[r0r]}$  in 4D $\cN=4$ with $\mathrm{R}$-symmetry $[r,0,r]$ and dimension $2r$ has the following number of superconformal invariants:
			\begin{equation}\label{eq:qbpscounting}
				\dim\left( \langle \cO_{[r0r]} \cO_{[r0r]} \cO_{[r0r]}\rangle\right) = \binom{r+2}{r}.
			\end{equation}
			Three-point functions involving quarter-BPS operators were studied in \cite{DHoker:2001jzy}.
			A general classification of such three-point structures together with the superconformal weight-shifting operators of \cite{Hansen:2025lig} give all the necessary tools to derive specific non-half-BPS superconformal blocks, i.e. contributions of an entire superconformal multiplet to a four-point function.
			This remains an unsolved problem in general, whose solution directly advances the superconformal bootstrap. 
			In the same context, it would be interesting to see whether the analyticity requirements on analytic superspace can be used to simplify or even evaluate shadow integrals \cite{Simmons-Duffin:2012juh}, and thus (super)conformal blocks, directly. Shadow integrals in this context were constructed in \cite{Hansen:2025lig} but have not been explored very much as yet.

			A new SUSY-specific feature of three-point superconformal invariants of non-half-BPS multiplets arises from requiring only $\SU(m,m|2n)$ covariance rather than $\U(m,m|2n)$ invariance. At two points the former implies the latter, but at three points there are non-trivial solutions of $\SU(m,m|2n)$ Ward identities which do not solve $\U(m,m|2n)$ Ward identities~\cite{Heslop:2003xu}.  Such three-point structures contain a novel constant supersymmetric $\mathcal E$-tensor which is invariant under $\SL(m|n)$ but not $\GL(m|n)$. 
			These then directly relate to the appearance of five-point nilpotent invariants, which vanish if the Grassmann odd components $\rho, \bar{\rho}$ of the superspace variable \eqref{eq:analyticX} are set to $0$. 
			The counting \eqref{eq:qbpscounting} will miss  such invariants.

			The universal form of three-point structures in $\SU(m,m|2n)$ analytic superspace makes clear the direct relation between  finite-dimensional $\SL(2n)$ representation theory 
			and the conformal group $\SU(2,2)$. The tensor structures are defined by universal $SU(m,m|2n)$ Ward identities, together with conservation conditions.  Both of these are differential equations  where only the dimension of the indices changes with $m,n$ and the solutions (and counting) therefore have a universal form.
			Similar observations could be made for more general correlation functions. 
			This reveals  a tensor product preserving duality between the two groups. 
			This is particularly clear in the counting of conformally invariant tensor structures from Littlewood-Richardson coefficients performed in section \ref{sec:LRcounting} (with further details in appendix \ref{app:LR}). 
			It would be interesting to understand and/or place this  within a more general mathematical framework.  
			One highly suggestive such framework is the theta correspondence (or Howe duality conjecture) \cite{Howe1989} which has not been very widely discussed in physics, although possible applications to physics have been reviewed in \cite{Basile:2020gqi} and it would be fascinating to see if the universality of analytic superspace  could be viewed from that perspective.

			\section*{Acknowledgements}
			We would like to thank Francesco Aprile, Tobias Hansen, Zhongjie Huang, Paul McFadden and Antoine Vauterin for helpful discussions. HPR is supported by an STFC studentship. PH is supported by STFC Consolidated Grant ST/X000591/1.
			
			\begin{appendix}
				\numberwithin{equation}{section}

				\section{Conformal frame, identities \& checks}\label{app:CF}

				We have checked our main formula for the basis of conserved tensor structures~\eqref{main}  systematically in low-spin cases generated with the CFTs4D \textsf{Mathematica} package of \cite{Cuomo:2017wme}. This can be done efficiently
				by fixing a conformal frame. For this, we used the conformal frame defined in the package, given by
				\begin{equation}
					\begin{aligned}
						x_1^\mu &= (0,0,0,0) &&\qquad & x_{1}^{\alpha \dota}&= \begin{pmatrix}
							0&0\\0&0
						\end{pmatrix} \\
						x_2^\mu &= \left((\bar{z}-z)/2,0,0,(z+\bar{z})/2\right) && \qquad\leftrightarrow\qquad &x_2^{\alpha \dota}&=\begin{pmatrix}
							\bar{z} & 0 \\
							0 &-z
						\end{pmatrix}\\
						x_3^\mu &= (0,0,0,1) &&\qquad & x_3^{\alpha \dota}&=\begin{pmatrix}
							1 & 0 \\
							0 & -1
						\end{pmatrix}
					\end{aligned} 
				\end{equation}
				and contracted the Weyl spinor indices at each point $X_i$ with polarisation spinors
				\begin{equation}\label{eq:polarisations}
					s^{\alpha}_i = \begin{pmatrix}
						\xi_i \\ \eta_i
					\end{pmatrix}, \qquad \qquad \bar{s}^{\dota}_i=\begin{pmatrix}
						\bar{\xi}_i & \bar{\eta}_i
					\end{pmatrix}.
				\end{equation}
				Doing so yields the following explicit form of the matrix of basic covariants \eqref{Vs}
				\begin{align}
					V_{ij} \  &=  \ \begin{pmatrix} Y_{1}&I_{12}&I_{13}\\-I_{21}&Y_{2}&I_{23}\\-I_{31}&-I_{32} &Y_{3}\end{pmatrix}\\ \notag
					&= \
					\begin{pmatrix}
						\dfrac{(z-1)\eta_1\bar\eta_1}{z}-\dfrac{(\bar z-1)\xi_1\bar\xi_1}{\bar z}
						&
						-\dfrac{\eta_2\bar\eta_1}{z}+\dfrac{\xi_2\bar\xi_1}{\bar z}
						&
						-\eta_3\bar\eta_1+\xi_3\bar\xi_1
						\\[0.9em]
						-\dfrac{\eta_1\bar\eta_2}{z}+\dfrac{\xi_1\bar\xi_2}{\bar z}
						&
						\dfrac{\eta_2\bar\eta_2}{(z-1)z}
						+
						\dfrac{\xi_2\bar\xi_2}{\bar{z}(1-\bar z)}
						&
						\dfrac{\eta_3\bar\eta_2}{z-1}-\dfrac{\xi_3\bar\xi_2}{\bar z-1}
						\\[0.9em]
						-\eta_1\bar\eta_3+\xi_1\bar\xi_3
						&
						\dfrac{\eta_2\bar\eta_3}{z-1}-\dfrac{\xi_2\bar\xi_3}{\bar z-1}
						&
						\dfrac{z\,\eta_3\bar\eta_3}{z-1}-\dfrac{\bar z\,\xi_3\bar\xi_3}{\bar z-1}
					\end{pmatrix}.
				\end{align}
				We note the difference in normalisation between $I_{ij}$ and $Y_i$ and the analogous covariants used in \cite{Cuomo:2017wme,Elkhidir:2014woa}, which include additional powers of $\det(x_{ij})$. Thus, for a direct comparison we need to include the scalar prefactor \eqref{eq:K3}.
				For example, our results for $\langle JJJ\rangle$ \eqref{eq:JJJ} and $\langle TTT \rangle$ \eqref{eq:TTT} as directly constructed from \eqref{main} written in the basis of the CFTs4D package with coefficients $\{\lambda_1^+, \dots, \lambda^-_1,\dots\}$, are given by
				\begin{equation}
					\begin{aligned}
						\llangle JJJ\rrangle^{(0)} &= \left\{ \lambda_1^+ = 1,\ \lambda_2^+= 1,\ \lambda_1^-=0\right\}, \\
						\llangle JJJ\rrangle^{(\pm 1)} &= \left\{ \lambda_1^+ = -\frac12,\ \lambda_2^+= \frac12,\ \lambda_1^-=\mp \frac12\right\},
					\end{aligned}
				\end{equation}
				and 
				\begin{equation}
					\begin{aligned}
						\llangle TTT \rrangle^{(0)} &= \left\{ \lambda_1^+ = \frac18,\ \lambda_2^+= \frac12,\ \lambda_3^+= \frac18,\ \lambda_1^-=0, \ \lambda_2^-=0\right\}, \\
						\llangle TTT \rrangle^{(\pm 1)} &= \left\{ \lambda_1^+ = \frac12,\ \lambda_2^+=- \frac14,\ \lambda_3^+= -\frac14,\ \lambda_1^-=\pm \frac12, \ \lambda_2^-=\pm \frac14\right\}, \\
						\llangle TTT \rrangle^{(\pm 2)} &= \left\{ \lambda_1^+ = \frac{1}{16},\ \lambda_2^+=- \frac18,\ \lambda_3^+= \frac{1}{16},\ \lambda_1^-=\pm  \frac{1}{16}, \ \lambda_2^-=\mp  \frac{1}{16}\right\}. 
					\end{aligned}
				\end{equation}
				
				The conformal frame is also used to map the more unconventional form of tensor structures derived in section \ref{sec:nonconserved} (non-conserved third insertion) to other existing bases.
				For instance, consider our result for $\langle J J J^2 \rangle$ \eqref{eq:JJphi6} obtained by constructing an explicitly pole-free term in the diagonal mean-field basis. First, we note that antisymmetrisation of Weyl spinor indices is equivalent to index contraction, and writing one of the inverses as $M^{-1}\sim \mathrm{adj}\,M/\det M$, we have
				\begin{equation}
					\begin{aligned}
						I_{13}{}_{\dota_1 [\alpha_3} I_{23}{}_{|\dota_2| \beta_3]} &= \frac{I_{13}{}_{\dota_1 \alpha_3 } X_{32}^{\alpha_3 \dotb_2} \epsilon_2{}_{\dotb_2 \dota_2}}{\det X_{32}} \\
						I_{31}{}_{[\dota_3| \alpha_1|} I_{32}{}_{\dotb_3] \alpha_2} &= \frac{\epsilon_1{}_{\alpha_1 \beta_1} X_{13}^{\beta_1\dota_3} I_{32}{}_{\dota_3 \alpha_2 }}{\det X_{13}}
					\end{aligned}
				\end{equation}
				This can now be contracted with the polarisations \eqref{eq:polarisations} and written in the conformal frame as
				\begin{equation}
					\begin{aligned}
						\bar{s}_1^{\dfa_1} I_{13}{}_{\dota_1 \alpha_3 } X_{32}^{\alpha_3 \dotb_2} \epsilon_2{}_{\dotb_2 \dota_2} \bar{s}_2^{\dfa_2}\ &= \ (\bar{z}-1) \bar{\eta}_2 \bar{\xi}_1 - (z-1) \bar{\eta}_1 \bar{\xi}_2, \\
						s_1^{\alpha_1 }\epsilon_1{}_{\alpha_1 \beta_1} X_{13}^{\beta_1\dota_3} I_{32}{}_{\dota_3 \alpha_2 } s_2^{\alpha_2}  \ &= \  \frac{\eta_2 \xi_1}{1-z} + \frac{\eta_1 \xi_2}{\bar{z}-1}.
					\end{aligned}
				\end{equation}
				Including the appropriate scalar prefactor, we find that \eqref{eq:JJphi6} is equal to \eqref{eq:JJphi6v2}
				\begin{equation}
					\frac{x_{12}^2}{(x^2_{13})^3 (x_{23}^2)^3}( I_{12} I_{21} + Y_1 Y_2),
				\end{equation}
				also in agreement with the CFTs4D package.

				\section{Computing Littlewood-Richardson coefficients}\label{app:LR}
				Littlewood-Richardson (LR) coefficients $C^{\ula}_{\umu \unu}$ count the number of different Young tableaux\footnote{A Young tableau is a Young diagram filled with an alphabet, usually integers.} of shape $\ula$ that can be constructed from a `product' of two other tableaux of shapes $\umu, \unu$. This number will be zero unless
				\begin{equation}\label{eq:LRreqs}
					|\ula|=|\umu|+|\unu|, \qquad \qquad \umu,\unu \subseteq \ula.
				\end{equation}
				See \cite{Fulton_yts, Fulton_reps} for a review of the combinatorics and $S_N$ representation theory of LR coefficients.
				
				By Schur-Weyl duality, finite-dimensional $\GL(n)$ representations are labelled by Young diagrams, and their tensor products are then also described by the Littlewood-Richardson rule. Indeed, let $V=\mathbb{C}^n$ be the fundamental of $\GL(n)$, then all finite-dimensional representations are constructed by applying a Young symmetriser $\Pi_{\ula}$ to $|\ula|$ tensor copies of $V$ as follows
				\begin{equation}
					V_{\ula} := \Pi_{\ula}\left( V^{\otimes|\ula|}\right).
				\end{equation}
				For example, the symmetric and alternating representations from two copies of $V$:
				\Yboxdim{6pt}
				\begin{equation}
					V_{\yng(2)}=\Pi_{\yng(2)}\left( V^{\otimes2}\right) \equiv \mathrm{Sym}^2V, \qquad \qquad V_{\yng(1,1)}=\Pi_{\yng(1,1)}\left( V^{\otimes2}\right) \equiv \wedge^2V\, .
				\end{equation}
				The Littlewood-Richardson rule for such $\GL(n)$ representations is 
				\begin{equation}
					V_{\umu} \otimes V_{\unu} \cong \bigoplus_{\ula} (V_{\ula})^{\oplus C^{\ula}_{\umu \unu}}\ .
				\end{equation}
				We also note that LR coefficients give the branching rules for $\GL(m+n)\to \GL(m)\times \GL(n)$ as follows. Let $W$ and $U$ be fundamentals of $\GL(m+n)$ and $\GL(m)$, respectively. Then, we have that
				\begin{equation}
					W_{\ula} |_{\GL(m)\times \GL(n)} = \bigoplus_{\umu,\unu} (V_{\umu} \otimes U_{\unu})^{\oplus C^{\ula}_{\umu \unu}}.
				\end{equation}
				The above equations can be interpreted in terms of Schur polynomials $s_{\ula}(x_1,\dots,x_n)$, which are characters of $\GL(n)$, i.e. traces of $V_{\ula}$. 
				
				To compute Littlewood-Richardson coefficients $C^{\ula}_{\umu \unu}$, one has to determine the number of semistandard LR fillings of the shape $\ula/\umu$ (or $\ula/\unu$, they are symmetric under $\umu \leftrightarrow \unu$) of weight $\unu$. We will now briefly define each of the terms.
				
				The skew-diagram $\ula/ \umu$ is obtained by removing $\umu$ from $\ula$ as follows
				\begin{equation}
					\ula \ = \  \begin{tikzpicture}[baseline=(align)]
						\draw[fill=gray!10, draw=none] 
						(2,0)--(3,0)--(3,-1)--(2.5,-1)--(2.5,-1.5)--(1.5,-1.5)--(1.5,-2.5)--(0.5,-2.5)--(0.5,-3)--(0,-3)--(0,-2.5)--(1,-2.5)--(1,-1.5)--(1.5,-1.5)--(1.5,-1)--(2,-1)--(2,0);
						\draw[dashed] (0,0)--(2,0)--(2,-1)--(1.5,-1)--(1.5,-1.5)--(1,-1.5)--(1,-2.5)--(0.5,-2.5)--(0,-2.5);
						\node at (0.75,-1) {$\umu$};
						\draw[thick] (0,0)--(3,0)--(3,-1)--(2.5,-1)--(2.5,-1.5)--(1.5,-1.5)--(1.5,-2.5)--(0.5,-2.5)--(0.5,-3)--(0,-3)--(0,0);
						\node (skew) at (2.5,-0.5) {$\ula/\umu$};
						\node (align) at (0,-1.25) {};
					\end{tikzpicture}\ .
				\end{equation}
				A filling of $\ula/\unu$ is said to be of weight $\unu$ if it contains the following set of integers
				\begin{equation}
					\{ i\in \mathbb{N} \ : \ i \text{ appears } \nu_i \text{ times}\},
				\end{equation}
				for $\unu=[\nu_1,\nu_2,\dots]$.
				Moreover, a filling is said to be semistandard if the integers are arranged in weakly increasing order along rows and strictly increasing down columns. Finally, this filling is a LR filling if it satisfies the Lattice word condition. That is,
				when reading the entries from right to left and top to bottom, any given integer $i$ must satisfy
				\begin{equation}
					\#(i) \geq \#(i+1),
				\end{equation}
				where $\#$ counts the number of occurrences in the specified order.

				\subsection{Conserved symmetric traceless tensors}\label{sec:current}
				A symmetric traceless tensor satisfies $s=\bar{s}$ and thus is labelled by a single spin label $J_s$. A correlator of three such conserved currents in 4D was conjectured \cite{Costa:2011mg} to contain
				\begin{equation}\label{eq:numstruct1}
					\dim \left( \langle J_{s_1} J_{s_2}J_{s_3} \rangle \right) = 2 \min (s_1,s_2,s_3)+1
				\end{equation}
				tensor structures. This was checked in several cases by \cite{Stanev:2012nq,Cuomo:2017wme,Buchbinder:2023coi}. We showed and generalised this formula  analytically in the main text \eqref{eq:counting}. We now provide an alternative proof via the Littlewood-Richardson rule by following the process outlined in section \ref{sec:LRcounting}.
				
				The first step is to lift $J_s$ to an $(m,n)$ superfield \eqref{eq:conservedsuperfield} and then consider the $m=0$ version which is more simply written as a $\SL(2n)$ Young diagram \eqref{eq:SL2nYD} with $s=\bar{s}$:
				\begin{equation}\label{eq:conserveddual}
					\uLa_{i}:=\uLa_{s_i,s_i} = [2^{n-s_i},1^{2s_i}] \equiv \begin{tikzpicture}[x=14pt, y=16pt,scale=1.3, baseline=(align)]
						\node (align) at (0.25,-1.7) {};
						\draw [ line width=0.6pt ] (0,0) rectangle (0.5,-2);
						\draw [ line width=0.6pt ] (0,-2) rectangle (0.5,-2.9);
						\draw [ line width=0.6pt ] (0.5,0) rectangle (1,-1.1);
						\node at (0.25,-0.82) {$\vdots$};
						\draw [line width=0.6pt,<->] (-0.25,0) -- (-0.25,-1.95)node[midway,left]{$n$};]
						\draw [line width=0.6pt,<->] (-0.25,-2.05) -- (-0.25,-2.9)node[midway,left]{$s_i$};
						\draw [line width=0.6pt,<->] (1.25,-1.1) -- (1.25,-2)node[midway,right]{$s_i$};
						\draw [line width=0.6pt,dashed] (0.5,-2) -- (1.5,-2);
					\end{tikzpicture}\, ,
				\end{equation}
				Let $[1^{2n},\uLa]$ be the Young diagram obtained from $\uLa$ by adding a column of length $2n$ to the left of $\ula$ ($c=1$ in \eqref{eq:SL2nYD}). Then, it suffices to show that
				\begin{equation}\label{eq:LRconjecture1}
					C^{[1^{2n},\uLa_3]}_{\uLa_1, \uLa_2}= 2 \min(s_1, s_2, s_3)+1,
				\end{equation}
				for $n$ above the stability threshold to be determined.
				
				Let us choose the ordering $s_1 \leq s_2 \leq s_3$ without loss of generality, since the proof for the other cases follows from the symmetry of the LR coefficients.
				By the Littlewood-Richardson rule, the coefficient \eqref{eq:LRconjecture1} corresponds to the number of inequivalent ways of filling the skew-diagram
				\begin{equation}\label{eq:skewdiagram1}
					[1^{2n},\uLa_3]/ \uLa_2 = \begin{tikzpicture}[x=14pt, y=14pt,scale=1.3, baseline=(align)]
						\node (align) at (0.25,-2) {};
						\draw [ line width=0.6pt ] (-0.5,0) rectangle (0,-4);
						\draw [line width=0.6pt, fill=black] (-0.5,0) rectangle (0,-2.7);
						\draw [line width=0.6pt, fill=black] (0,0) rectangle (0.5,-1.3);
						\draw [ line width=0.6pt ] (0,0) rectangle (0.5,-3);
						\draw [ line width=0.6pt ] (0.5,0) rectangle (1,-1);
						\draw [line width=0.6pt,<->] (-0.75,-2.05) -- (-0.75,-2.7)node[midway,left]{$s_2$};
						\draw [line width=0.6pt,<->] (1.25,-1) -- (1.25,-2)node[midway,right]{$s_3$};
						\draw [line width=0.6pt,dashed] (-1,-2) -- (1.3,-2);
					\end{tikzpicture},
				\end{equation}
				with the following set of integers
				\begin{equation}\label{eq:sssLRintegers}
					\{ \underbrace{1,1,2,2, \dots, n-s_1, n-s_1}_{\times 2}, n-s_1+1,\dots, n+s_1\},
				\end{equation}
				such that the rows are weakly increasing, columns strictly increasing and any word read from top to bottom and right to left satisfies the lattice word condition.
				
				These conditions fully constrain the first column to be filled by $\{1,\dots, n-s_3\}$ in order.
				Moreover, since multiplicity-1 integers are strictly larger than those with multiplicity 2, they must appear at the ends of columns 2 or 3 in strictly increasing order.
				Let $\mathbf{k}^i_1$ be a column tableau filled from top to bottom by $\{n-s_1+1, \dots, n-s_1+i\}$ and 
				$\mathbf{k}^i_2$ by $\{n-s_1+i+1, \dots, n+s_1\}$, for $i\in [0,2s_1]$. Then, all valid fillings of \eqref{eq:skewdiagram1} are of the form
				\begin{equation}\label{eq:filling11}
					\begin{tikzpicture}[x=20pt, y=20pt,scale=1.4, baseline=(align)]
						\node (align) at (0.25,-2) {};
						\node at (0.75, -0.2) {$\scriptstyle{1}$};
						\node at (0.75, -0.5) {$\scriptstyle{2}$};
						\node at (0.75, -0.72) {$\scriptstyle{\cdot}$};
						\node at (0.75, -0.82) {$\scriptstyle{\cdot}$};
						\node at (0.75, -0.92) {$\scriptstyle{\cdot}$};
						\draw [line width=0.6pt,<->] (1.25,0) -- (1.25,-1)node[midway,right]{$n\!-\!s_3$};
						\draw [line width=0.6pt,<->] (1.25,-1.1) -- (1.25,-3)node[midway,right]{$s_2\!+\!s_3$};
						\draw [ line width=0.6pt ] (0,-1.1) rectangle (0.5,-3);
						\node at (0.25, -1.3) {$\scriptstyle{1}$};
						\node at (0.25, -1.55) {$\scriptstyle{\cdot}$};
						\node at (0.25, -1.65) {$\scriptstyle{\cdot}$};
						\node at (0.25, -1.75) {$\scriptstyle{\cdot}$};
						\draw [ line width=0.6pt ] (0,-2.2) -- (0.5,-2.2);
						\node at (0.26, -2.6) {$\mathbf{k}^i_1$};
						\draw [ line width=0.6pt ] (0.5,0) rectangle (1,-1);
						\draw [ line width=0.6pt ] (-0.5,-2.7) rectangle (0,-4);
						\draw [ line width=0.6pt ] (-0.5,-3.35) -- (0,-3.35);
						\node at (-0.24, -3.7) {$\mathbf{k}^i_2$};
						\draw [line width=0.6pt,dashed] (-0.5,-2) -- (1,-2);
					\end{tikzpicture}
					.
				\end{equation}
				It then follows that for any given choice of $i\in [0,2s_1]$ there exists at most 1 filling of the remaining multiplicity 2 integers 
				$$\{1,\dots,n-s_3,\underbrace{n-s_3+1,\dots, n-s_1}_{\times 2}\}$$ 
				fulfilling strictly increasing and lattice word conditions. 
				Thus, the number of possible fillings is at most $$|[0,2 s_1]|=2s_1 +1.$$
				This bound is saturated when $n-s_1 \geq s_2+s_3$, i.e. there are at least as many multiplicity-2 integers as boxes in the second column. 
				This determines the stability threshold $k$ of the Littlewood-Richardson coefficients as
				\begin{equation}
					n \geq k=s_1 + s_2 + s_3 .
				\end{equation}
				in agreement with \eqref{eq:counting2}.
				This completes the proof of \eqref{eq:numstruct1}. 
				
				For example, consider three currents ($s_1=s_2=s_3=1$) at the stability threshold $n=3$ which translates to
				\Yboxdim{8pt}
				\begin{equation}\label{eq:skews1}
					\Yvcentermath1
					\ula_1 =\ula_2=\ula_3= \yng(2,2,1,1)\,, \qquad [1^{2n},\ula_3]= \yng(3,3,2,2,1,1) \qquad \Rightarrow \qquad [1^{2n},\ula_3]/\ula_2 = \begin{tikzpicture}[x=16pt, y=16pt,scale=1, baseline=(align)]
						\node (align) at (0.25,-1.5) {};
						\tyng(3,3,3,3,2,2,1,1)
						\draw [fill=black] (0,0.5) rectangle (1,-0.5);
						\draw [fill=black] (0.5,-1.5) rectangle (0,-0.5);
					\end{tikzpicture}\,,
				\end{equation}
				and the set of integers $\{ 1,1,2,2,3,4\}$.
				The three semistandard LR fillings of \eqref{eq:skews1} are given by
				\begin{equation}\label{eq:JJJfillings}
					\begin{tikzpicture}[x=14pt, y=14pt,scale=1.1, baseline=(align)]
						\node (align) at (0.25,-2.4) {};
						\pgfmathsetmacro{\le}{0.7}
						\node at (0,-1*\le) {$\scriptstyle 1$};
						\node at (0,-2*\le) {$\scriptstyle 2$};
						\node at (-1*\le,-3*\le) {$\scriptstyle \mathbf{3}$};
						\node at (-1*\le,-4*\le) {$\scriptstyle \mathbf{4}$};
						\node at (-2*\le,-5*\le) {$\scriptstyle 1$};
						\node at (-2*\le,-6*\le) {$\scriptstyle 2$};
						\draw [line width=0.6] (0.5*\le,-0.5*\le) rectangle (-.5*\le,-2.5*\le);
						\draw [line width=0.6] (-1.5*\le,-4.5*\le) rectangle (-.5*\le,-2.5*\le);
						\draw [line width=0.6] (-1.5*\le,-4.5*\le) rectangle (-2.5*\le,-6.5*\le);
					\end{tikzpicture}\,, \qquad \quad     \begin{tikzpicture}[x=14pt, y=14pt,scale=1.1, baseline=(align)]
						\node (align) at (0.25,-2.4) {};
						\pgfmathsetmacro{\le}{0.7}
						\node at (0,-1*\le) {$\scriptstyle 1$};
						\node at (0,-2*\le) {$\scriptstyle 2$};
						\node at (-1*\le,-3*\le) {$\scriptstyle 1$};
						\node at (-1*\le,-4*\le) {$\scriptstyle \mathbf{3}$};
						\node at (-2*\le,-5*\le) {$\scriptstyle 2$};
						\node at (-2*\le,-6*\le) {$\scriptstyle \mathbf{4}$};
						\draw [line width=0.6] (0.5*\le,-0.5*\le) rectangle (-.5*\le,-2.5*\le);
						\draw [line width=0.6] (-1.5*\le,-4.5*\le) rectangle (-.5*\le,-2.5*\le);
						\draw [line width=0.6] (-1.5*\le,-4.5*\le) rectangle (-2.5*\le,-6.5*\le);
					\end{tikzpicture}\,, \qquad \quad    \begin{tikzpicture}[x=14pt, y=14pt,scale=1.1, baseline=(align)]
						\node (align) at (0.25,-2.4) {};
						\pgfmathsetmacro{\le}{0.7}
						\node at (0,-1*\le) {$\scriptstyle 1$};
						\node at (0,-2*\le) {$\scriptstyle 2$};
						\node at (-1*\le,-3*\le) {$\scriptstyle 1$};
						\node at (-1*\le,-4*\le) {$\scriptstyle 2$};
						\node at (-2*\le,-5*\le) {$\scriptstyle \mathbf{3}$};
						\node at (-2*\le,-6*\le) {$\scriptstyle \mathbf{4}$};
						\draw [line width=0.6] (0.5*\le,-0.5*\le) rectangle (-.5*\le,-2.5*\le);
						\draw [line width=0.6] (-1.5*\le,-4.5*\le) rectangle (-.5*\le,-2.5*\le);
						\draw [line width=0.6] (-1.5*\le,-4.5*\le) rectangle (-2.5*\le,-6.5*\le);
					\end{tikzpicture}\,,
				\end{equation}
				in agreement with \eqref{eq:numstruct1} and \eqref{eq:filling11}.
				This number and structure remains stable for all $n\geq3$ matching the number of tensor structures of a correlation function of three conserved currents (2 parity even and 1 parity odd).
				
				\subsection{Mixed symmetry tensors}\label{sec:mixed}
				The next simplest family of correlators present in the general counting formula \eqref{eq:counting} is that consisting of two mixed symmetry representations and a third symmetric traceless insertion
				\begin{equation}
					\langle J_{s_1,\bar{s}_1}(x_1) J_{s_2,\bar{s}_2}(x_2)J_{s_3}(x_3)\rangle, \qquad \quad s_1+s_2=\bar{s}_1+\bar{s}_2.
				\end{equation}
				In \cite{Buchbinder:2023coi}, a subfamily of these satisfying
				\begin{equation}\label{eq:numstruct2}
					\dim \left( \langle J_{s_1+p,s_1-p} J_{s_2-p,s_2+p}J_{s_3}\rangle\right) =    1+2\min\left(s_1,s_2,s_3\right) - \max\left(0,p-|s_3-\min(s_1,s_2)|\right).
				\end{equation}
				It is straightforward to check that this agrees with our general formula \eqref{eq:counting}. We will now provide an alternative analytic derivation via the LR rule.
				
				Let $\uGa_i$ be the $\SL(2n)$ Young diagram obtained from lifting $J_{s_i+p, s_i-p}$ to a $(m,n)$ superfield \eqref{eq:conservedsuperfield} and restricting to $m=0$. Let $\tilde{\uGa}_i$ be the corresponding lift and restriction of $J_{s_i-p, s_i+p}$. They are given by
				\begin{equation}
					\uGa_i = \begin{tikzpicture}[x=14pt, y=17pt,scale=1.2, baseline=(align)]
						\node (align) at (0.25,-1.7) {};
						\draw [ line width=0.6pt ] (0,0) rectangle (0.5,-2);
						\draw [ line width=0.6pt ] (0,-2) rectangle (0.5,-2.9);
						\draw [ line width=0.6pt ] (0.5,0) rectangle (1,-0.7);
						\node at (0.25,-0.82) {$\vdots$};
						\draw [line width=0.6pt,<->] (-0.25,0) -- (-0.25,-1.95)node[midway,left]{$n$};]
						\draw [line width=0.6pt,<->] (-0.25,-2.05) -- (-0.25,-2.9)node[midway,left]{$s_i\!-\!p$};
						\draw [line width=0.6pt,<->] (1.25,-0.7) -- (1.25,-2)node[midway,right]{$s_i\!+\!p$};
						\draw [line width=0.6pt,dashed] (0.5,-2) -- (1.5,-2);
					\end{tikzpicture} , \qquad \qquad \tilde{\uGa}_i =  \begin{tikzpicture}[x=14pt, y=17pt,scale=1.2, baseline=(align)]
						\node (align) at (0.25,-1.9) {};
						\draw [ line width=0.6pt ] (0,0) rectangle (0.5,-2);
						\draw [ line width=0.6pt ] (0,-2) rectangle (0.5,-3.3);
						\draw [ line width=0.6pt ] (0.5,0) rectangle (1,-1.1);
						\node at (0.25,-0.82) {$\vdots$};
						\draw [line width=0.6pt,<->] (-0.25,0) -- (-0.25,-1.95)node[midway,left]{$n$};]
						\draw [line width=0.6pt,<->] (-0.25,-2.05) -- (-0.25,-3.3)node[midway,left]{$s_i\!+\!p$};
						\draw [line width=0.6pt,<->] (1.25,-1.1) -- (1.25,-2)node[midway,right]{$s_i\!-\!p$};
						\draw [line width=0.6pt,dashed] (0.5,-2) -- (1.5,-2);
					\end{tikzpicture}.
				\end{equation}
				Clearly, the representation $[1^{2n},\uLa_3]$ is in the tensor product $\uGa_1\otimes \tilde{\uGa}_2$. It remains to find the multiplicity, i.e. show thay 
				\begin{equation}
					C^{[1^{2n},\uLa_3]}_{\uGa_1,\tilde{\uGa}_2}= \eqref{eq:numstruct2},
				\end{equation}
				for $n$ above a stability threshold $k$ (which we already know is given by \eqref{eq:counting2}).
				
				Note that by introducing mixed symmetry in the tensors at points 1 and 2, we have broken the symmetry that was previously used to fix an ordering of the spins without loss of generality. There are now two separate cases to consider for the full proof, namely $s_1 \leq s_2\leq s_3$ and $s_3\leq s_1 \leq s_2$. We consider the former here as the latter follows analogously.
				
				By the LR rule, the different tensor structures correspond the the different valid LR fillings of
				\begin{equation}\label{eq:skewdiagram2}
					[1^{2n},\ula_3]/ \tilde{\umu}_2 \ = \  \begin{tikzpicture}[x=14pt, y=14pt,scale=1.3, baseline=(align)]
						\node (align) at (0.25,-2) {};
						\draw [ line width=0.6pt ] (-0.5,0.2) rectangle (0,-4.2);
						\draw [line width=0.6pt, fill=black] (-0.5,0.2) rectangle (0,-2.9);
						\draw [line width=0.6pt, fill=black] (0,0.2) rectangle (0.5,-1.6);
						\draw [ line width=0.6pt ] (0,0.2) rectangle (0.5,-2.6);
						\draw [ line width=0.6pt ] (0.5,0.2) rectangle (1,-1.3);
						\draw [line width=0.6pt,<->] (0.25,-4.2) -- (0.25,-2.9)node[midway,right]{$n\!-\!s_2\!-\!p$};
						\draw [line width=0.6pt,<->] (0.75,-2.6) -- (0.75,-1.65)node[midway,right]{$s_3\!+\!s_2\!-\!p$};
						\draw [line width=0.6pt,<->] (1.25,0.2) -- (1.25,-1.3)node[midway,right]{$n\!-\!s_3$};
						\draw [line width=0.6pt,dashed] (-1,-2) -- (0.75,-2);
					\end{tikzpicture},
				\end{equation}
				with the following set of integers
				\begin{equation}
					\{ \underbrace{1,1,2,2, \dots, n-s_1-p, n-s_1-p}_{\times 2}, n-s_1-p+1,\dots, n+s_1-p\}.
				\end{equation}
				
				As in the previous section, the stability threshold $k$ follows from the requirement that the number of multiplicity-2 integers is greater or equal than the length of the middle column.
				Thus, in the present case this is
				\begin{equation}
					n-s_1-p \geq s_3+s_2-p \quad \Rightarrow \quad n\geq k=s_1+s_2+s_3,
				\end{equation}
				giving the same value for $k$.
				
				It is straightforward to see that if the number of multiplicity-2 integers is greater than the number of boxes in the first column, 
				i.e. 
				$s_3\geq s_1+q$, then the counting is identical to that of the previous section, giving $2s_1+1$ different fillings, in agreement with \eqref{eq:numstruct2}, namely
				\begin{equation}
					C^{[1^{2n},\uLa_3]}_{\uGa_1,\tilde{\uGa}_2}=2s_1+1, \qquad s_1\leq s_2\leq s_1+p \leq s_3.
				\end{equation}
				
				It remains to investigate the case $s_3\leq s_1+p$, where the construction of the LR fillings deviates. This is because there are not enough multiplicity-2 integers to fill the first column, and thus we must use some of the multiplicity-1 ones. This reduces the number of integers available to sort into two columns $\mathbf{k}_1^i,\mathbf{k}_2^i$ as in Eq.~\eqref{eq:filling11}. In particular, it reduces them by 
				\begin{equation}
					(n-s_3)-(n-s_1-p)= s_1 + p- s_3.
				\end{equation}
				This gives the expected Littlewood-Richardson coefficient
				\begin{equation}
					C^{[1^{2n},\uLa_3]}_{\uGa_1,\tilde{\uGa}_2}=2s_1+1-(s_1 + p - s_3)=s_1 +s_3 +1 -p, \qquad s_1\leq s_2\leq s_3\leq s_1+p.
				\end{equation}
				The concrete fillings can be constructed as follows.
				Let $\mathbf{a}$ be the column Young tableau filled with multiplicity-2 integers $\left\{1,2,\dots, n-s_1-p \right\}$ and $\mathbf{b}$ be similarly filled by the first $s_1 + p - s_3$ multiplicity-1 integers $\left\{n-s_1-p +1, \dots, n-s_3\right\}$. Then, all LR fillings of \eqref{eq:skewdiagram2} are of the following form
				
				\begin{equation}
					\begin{tikzpicture}[x=21pt, y=17pt,scale=1.3, baseline=(align)]
						\node (align) at (0.25,-2) {};
						\node at (0.75, 0) {$\mathbf{a}$};
						\draw [ line width=0.6pt ] (0.5,-0.45) -- (1,-0.45);
						\node at (0.75, -0.7) {$\mathbf{b}$};
						\draw [line width=0.6pt,<->] (1.25,0.5) -- (1.25,-1)node[midway,right]{$n\!-\!s_3$};
						\draw [line width=0.6pt,<->] (0.25,0.5) -- (0.25,-0.44)node[midway,left]{$n\!-\!s_1\!-\!p$};
						\draw [line width=0.6pt,<->] (0.75,-1.1) -- (0.75,-3)node[midway,right]{$s_2\!+\!s_3\!-\!p$};
						\draw [ line width=0.6pt ] (0,-1.1) rectangle (0.5,-3);
						\node at (0.25, -1.35) {$\scriptstyle{1}$};
						\node at (0.25, -1.65) {$\cdot$};
						\node at (0.25, -1.8) {$\cdot$};
						\node at (0.25, -1.95) {$\cdot$};
						\draw [ line width=0.6pt ] (0,-2.3) -- (0.5,-2.3);
						\node at (0.26, -2.65) {$\mathbf{k}^i_1$};
						\draw [ line width=0.6pt ] (0.5,0.5) rectangle (1,-1);
						\draw [ line width=0.6pt ] (-0.5,-2.7) rectangle (0,-4.5);
						\draw [ line width=0.6pt ] (-0.5,-3.5) -- (0,-3.5);
						\node at (-0.24, -4) {$\mathbf{k}^i_2$};
						\draw [line width=0.6pt,dashed] (-0.5,-2) -- (0.72,-2);
					\end{tikzpicture},
				\end{equation}
				where $\mathbf{k}_1^i$ and $\mathbf{k}_2^i$ are the column diagrams filled with the remaining multiplicity-1 integers $\{ n-s_3+1,\dots, n-s_3+i\}$ and $\{ n-s_3+i,\dots, n+s_1-p\}$, respectively. 
				
				\subsection{Non-conserved insertion}
				We consider the counting of tensor structures in
				\begin{equation}\label{eq:JJOs}
					\langle J_s(x_1) J_s(x_2) \cO_s^{\Delta}(x_3)\rangle,
				\end{equation}
				for dimension $\Delta$ above the unitarity bound.
				As explained in the main text, $\cO_s^{\Delta}$ should be written as the field that diagonalises the free field, i.e. the free-theory combination of two copies of $J_s$. The first such non-conserved field in the free $J_s\times J_s$ OPE  is
				\begin{equation}
					(\cO^{\Delta=4+3s}_{s})^{\nu_1 \dots \nu_s}=J_s^{\mu_1 \dots \mu_s} \partial_{\mu_1}\dots \partial_{\mu_s} J_s^{\nu_1 \dots \nu_s}.
				\end{equation}
				As an $(m,n)$ coset field this is
				\begin{equation}\label{eq:OsLR}
					\cO^{\gamma=4}_{[2s,s],[2s,s]} \  \overset{m=0}{\longmapsto} \  \begin{tikzpicture}[x=14pt, y=20pt,scale=1.2, baseline=(align)]
						\node (align) at (0.25,-1.7) {};
						\draw [ line width=0.6pt ] (0,0) rectangle (1,-2);
						\draw [ line width=0.6pt ] (1.5,-1.4) rectangle (1,0);
						\draw [ line width=0.6pt ] (2,-0.8) rectangle (1.5,0);
						\draw [line width=0.6pt,<->] (2.05,-1.41) -- (2.05,-2)node[midway,right]{$s$};
						\draw [line width=0.6pt,<->] (2.05,-1.39) -- (2.05,-0.8)node[midway,right]{$s$};
						\draw [ line width=0.6pt ] (0,-3.2) rectangle (0.5,-2);
						\draw [ line width=0.6pt ] (0.5,-2.6) rectangle (1,-2);
						\draw [line width=0.6pt,<->] (1.25,-2.59) -- (1.25,-2)node[midway,right]{$s$};
						\draw [line width=0.6pt,<->] (1.25,-2.61) -- (1.25,-3.2)node[midway,right]{$s$};
						
						\node at (0.5,-0.82) {$\vdots$};
						\draw [line width=0.6pt,<->] (-0.25,0) -- (-0.25,-1.95)node[midway,left]{$n$};]
						\draw [line width=0.6pt,dashed] (-0.5,-2) -- (2.2,-2);
						\draw [line width=0.6pt,dotted] (1.5,-1.4) -- (2.1,-1.4);
						\draw [line width=0.6pt,dotted] (0.5,-3.2) -- (1.25,-3.2);
					\end{tikzpicture}:= \underline{\Xi}_{\,s} \ ,
				\end{equation}
				where we wrote the corresponding finite-dimensional representation obtained after setting $m=0$. 
				
				We now have $|\underline{\Xi}_{\, s}|= 2|\uLa_s|$ and thus no padding is necessary to compute the LR coefficient
				$ C^{\underline{\Xi}_{\, s}}_{\uLa_s\uLa_s}$. To do so, we have to fill the skew diagram
				\begin{equation}
					\underline{\Xi}_{\, s}/\uLa_s \ = \ \begin{tikzpicture}[x=14pt, y=20pt,scale=1.2, baseline=(align)]
						\node (align) at (0.25,-1.7) {};
						\draw [ line width=0.6pt ] (0,0) rectangle (1,-2);
						\draw [ line width=0.6pt ] (1.5,-1.4) rectangle (1,0);
						\draw [ line width=0.6pt ] (2,-0.8) rectangle (1.5,0);
						\draw [ line width=0.6pt ] (0,-3.2) rectangle (0.5,-2);
						\draw [ line width=0.6pt ] (0.5,-2.6) rectangle (1,-2);
						\draw [fill=black] (0,0) rectangle (0.5,-2.6);
						\draw [fill=black] (0.5,0) rectangle (1,-1.4);
						\draw [line width=0.6pt,dashed] (-0.5,-2) -- (2.2,-2);
					\end{tikzpicture},
				\end{equation}
				with the set of integers \eqref{eq:sssLRintegers} with $s_1=s$.
				Starting from the right, the first two columns in any semistandard LR filling of the above are always fixed to contain the first $n-2s$ and $n-s$ multiplicity-2 integers, respectively.
				Filling the remaining entries as in \eqref{eq:filling11}, we find that the number of possible fillings is now restricted by the length of the last column: $s$ (whereas in the fully conserved cases above this was arbitrarily large). This gives a final counting of
				\begin{equation}\label{eq:JJOsLR}
					C^{\underline{\Xi}_{\, s}}_{\uLa_s\uLa_s}= s+1,
				\end{equation}
				which is stable for all $n$ as long as $\underline{\Xi}_s$ exists, since the length of the middle column equals the number of multiplicity-2 integers.\footnote{This was the requirement that gave the stability threshold in the previous sections} This reproduces the counting from pole cancellation arguments \eqref{eq:JJOscount}.
				
				In CFT, we know that the number of tensor structures for \eqref{eq:JJOs} should not depend on $\Delta$ as long as it does not saturate the unitarity bound. So the above LR coefficient \eqref{eq:JJOsLR} should be independent of $\Delta$, i.e. invariant under extension of the two-row part in $\cO^{\gamma=4}_{\ula_L\ula_R}$ \eqref{eq:OsLR} (as long as the resulting field is in the $J_s\times J_s$ OPE). 
				This property is indeed satisfied by LR coefficients.
				To see why, consider a non-conserved field of larger dimension than \eqref{eq:OsLR}. Taking the $m=0$ Young diagram we have
				\begin{equation}
					\begin{tikzpicture}[x=14pt, y=20pt,scale=1.2, baseline=(align)]
						\node (align) at (0.25,-1.7) {};
						\draw [ line width=0.6pt ] (0,0) rectangle (1,-2);
						\draw [ line width=0.6pt ] (1.5,-1.1) rectangle (1,0);
						\draw [ line width=0.6pt ] (2,-0.7) rectangle (1.5,0);
						\draw [line width=0.6pt,<->] (2.05,-1.11) -- (2.05,-2)node[midway,right]{$\delta$};
						\draw [line width=0.6pt,<->] (2.05,-1.09) -- (2.05,-0.7)node[midway,right]{$s$};
						\draw [ line width=0.6pt ] (0,-3.3) rectangle (0.5,-2);
						\draw [ line width=0.6pt ] (0.5,-2.9) rectangle (1,-2);
						\draw [line width=0.6pt,<->] (1.25,-2.89) -- (1.25,-2)node[midway,right]{$\delta$};
						\draw [line width=0.6pt,<->] (1.25,-2.91) -- (1.25,-3.3)node[midway,right]{$s$};
						\node at (0.5,-0.82) {$\vdots$};
						\draw [line width=0.6pt,<->] (-0.25,0) -- (-0.25,-1.95)node[midway,left]{$n$};]
						\draw [line width=0.6pt,dashed] (-0.5,-2) -- (2.2,-2);
						\draw [line width=0.6pt,dotted] (1.5,-1.1) -- (2.1,-1.1);
						\draw [line width=0.6pt,dotted] (0.5,-3.3) -- (1.25,-3.3);
					\end{tikzpicture}.
				\end{equation}
				Crucially, after taking the skew diagram of the above with $\uLa_s$, the possible LR fillings are still constrained by the length of the final column $(\delta+s)-\delta=s$, independently of the value of $\delta$, ultimately giving the same counting: $s+1$.
				
			\end{appendix}

			\bibliography{ref.bib}

		\end{document}